\documentclass[traditabstract]{aa}
\usepackage{txfonts}
\usepackage{graphicx}
\usepackage{float}
\usepackage{natbib}
\usepackage{hyperref}
\usepackage[english]{babel}
\usepackage{color}

\title{Evolution of the dust emission of massive galaxies up to z=4 and constraints on their dominant mode of star formation}

\author{Matthieu~B{\'e}thermin\inst{1,2} \and Emanuele Daddi\inst{2} \and Georgios Magdis\inst{3} \and Claudia Lagos\inst{1} \and Mark Sargent\inst{4} \and Marcus Albrecht\inst{5} \and Herv\'e Aussel\inst{2} \and Frank Bertoldi\inst{5} \and V\'eronique Buat\inst{6} \and Maud Galametz\inst{1} \and S\'ebastien Heinis\inst{7} \and Olivier Ilbert\inst{6} \and Alexander Karim\inst{5} \and Anton Koekemoer\inst{8} \and Cedric Lacey\inst{9}  \and Emeric Le Floc'h\inst{2}  \and Felipe Navarrete\inst{5} \and Maurilio Pannella\inst{2}  \and Corentin Schreiber\inst{2}  \and Vernesa Smol\v{c}i\'c\inst{10} Myrto Symeonidis\inst{4,11} \and Marco Viero\inst{12}}
\institute{European Southern Observatory, Karl-Schwarzschild-Str. 2, 85748 Garching, Germany \email{matthieu.bethermin@eso.org} \and 
Laboratoire AIM-Paris-Saclay, CEA/DSM/Irfu - CNRS - Universit\'e Paris Diderot, CEA-Saclay, pt courrier 131, F-91191 Gif-sur-Yvette, France \and 
Department of Physics, University of Oxford, Keble Road, Oxford OX1 3RH, UK \and 
Astronomy Centre, Department of Physics and Astronomy, University of Sussex, Brighton, BN1 9QH, UK \and 
Argelander-Institut f\"ur Astronomie, Universit\"at Bonn, Auf dem H\"ugel 71, D-53121 Bonn, Germany \and
Aix-Marseille Universit\'e, CNRS, LAM (Laboratoire d'Astrophysique de Marseille) UMR7326, 13388 Marseille, France \and
Department of Astronomy, University of Maryland, College Park, MD 20742-2421, USA \and
Space Telescope Science Institute, 3700 San Martin Drive, Baltimore, MD 21218, USA \and
Institute for Computational Cosmology, Department of Physics, University of Durham, South Road, Durham DH1 3LE, UK \and
University of Zagreb, Physics Department, Bijeni\v{c}ka cesta 32, 10002 Zagreb, Croatia \and
Mullard Space Science Laboratory, University College London, Holmbury St Mary, Dorking, Surrey RH5 6NT, UK \and
California Institute of Technology, 1200 East California Boulevard, Pasadena, CA 91125, USA}

\date{Received 19 September 2014 / Accepted 11 November 2014} 

\abstract{We aim to measure the average dust and molecular gas content of massive star-forming galaxies ($\rm > 3 \times 10^{10}\,M_\odot$) up to z=4 in the COSMOS field to determine if the intense star formation observed at high redshift is induced by major mergers or is caused by large gas reservoirs. Firstly, we measured the evolution of the average spectral energy distributions as a function of redshift using a stacking analysis of \textit{Spitzer}, \textit{Herschel}, LABOCA, and AzTEC data for two samples of galaxies: normal star-forming objects and strong starbursts, as defined by their distance to the main sequence. We found that the mean intensity of the radiation field $\langle U \rangle$ heating the dust (strongly correlated with dust temperature) increases with increasing redshift up to z=4 in main-sequence galaxies. We can reproduce this evolution with simple models that account for the decrease in the gas metallicity with redshift. No evolution of $\langle U \rangle$ with redshift is found in strong starbursts. We then deduced the evolution of the molecular gas fraction (defined here as $\rm M_{\rm mol}/(M_{\rm mol}+M_\star)$) with redshift and found a similar, steeply increasing trend for both samples. At z$\sim$4, this fraction reaches $\sim$60\%. The average position of the main-sequence galaxies is on the locus of the local, normal star-forming disks in the integrated Schmidt-Kennicutt diagram (star formation rate versus mass of molecular gas), suggesting that the bulk of the star formation up to z=4 is dominated by secular processes.}


\titlerunning{Evolution of the dust emission of massive galaxies up to z=4}

\authorrunning{B\'ethermin et al.}

\begin{document}

\maketitle

\section{Introduction}

Galaxy properties evolve rapidly across cosmic time. In particular, various studies have shown that the mean star formation rate (SFR) at fixed stellar mass increases by a factor of about 20 between z=0 and z=2 \citep[e.g.,][]{Noeske2007,Elbaz2007,Daddi2007,Pannella2009,Magdis2010,Karim2011,Elbaz2011,Rodighiero2011,Whitaker2012,Heinis2014,Pannella2014}. This very high SFR can be explained by either larger reservoirs of molecular gas or a higher star formation efficiency (SFE). Large gas reservoirs have been found in massive galaxies at high redshift \citep[e.g.,][]{Daddi2008,Tacconi2010,Daddi2010a,Tacconi2013,Aravena2013}, which could imply high SFRs with SFE similar to that of normal star-forming galaxies in the local Universe. On the other hand, follow-up of bright submillimeter galaxies (SMGs) revealed that their very intense SFR ($\sim$1000\,M$_\odot$/yr) is also driven by a SFE boosted by a factor of 10 with respect to normal star-forming galaxies in the local Universe \citep[e.g.,][]{Greve2005,Frayer2008,Daddi2009a,Daddi2009b}, likely induced by a major merger. This difference can be understood if we consider that galaxies are driven by two types of star formation activity: smooth processes fed by large reservoirs of gas in normal star-forming galaxies and boosted star-formation in gas rich mergers \citep{Daddi2010b,Genzel2010}.\\

Using models based on the existence of this main-sequence of star-forming galaxies, i.e., a tight correlation between SFR and stellar mass, and outliers of this sequence with boosted sSFRs (SFR/M$_\star$) called starbursts hereafter, \citet{Sargent2012} showed that the galaxies with the highest SFR mainly correspond to starbursts, while the bulk of the star formation budget ($\sim$85\%) is hosted in normal star-forming galaxies. This approach allows us to better understand the heterogeneous characteristic of distant objects concerning their gas fraction and their SFE \citep{Sargent2014}. The quick rise of the sSFR would thus be explained by larger gas reservoirs in main-sequence galaxies. However, the most extreme SFRs observed in high-redshift starbursts would be caused by a SFE boosted induced by major mergers.\\
 
 At high redshift, the gas mass is difficult to estimate. Two main methods are used. The first is based on the measurement of the intensity of rotational transitions (generally with J$_{\rm upper}<3$) of $^{12}$CO and an assumed CO-to-H$_2$ conversion factor \citep{Daddi2008,Tacconi2010,Saintonge2013,Tacconi2013}. The main limitation of this method is the uncertainty on this conversion factor, which is expected to be different from the local calibrations in high-redshift galaxies with strongly sub-solar metallicities \citep{Bothwell2010,Engel2010,Genzel2012,Tan2013,Genzel2014}. The second method is based on the estimate of the dust mass, which is then converted into gas mass using the locally-calibrated relation between the gas-to-dust ratio and the gas metallicity \citep[e.g.,][]{Munoz-Mateos2009,Leroy2011,Remy2014}. The main weakness of this method is the need of an accurate estimate of the gas metallicity and the possible evolution in normalization and scatter of the relation between gas-to-dust ratio and gas metallicity. This method was applied on individual galaxies at high redshift by \citet{Magdis2011,Magdis2012b} and \citet{Scoville2014}, but also on mean spectral energy distributions (SEDs) measured through a stacking analysis \citep{Magdis2012b,Santini2014}. This method has not been applied at redshifts higher than $\sim$2. The aim of this paper is to extend the studies of dust emission and gas fractions derived from dust masses to z$\sim$4 and analyze possible differences in trends as redshift increases.\\

In this paper, we combine the information provided by the \textit{Herschel} data and a mass-selected sample of galaxies built from the UltraVISTA data \citep{Ilbert2013} in COSMOS to study the mean dust emission of galaxies up to z=4 (Sect.\,\ref{data}). We measure the mean SED of galaxies on the main sequence and strong starbursts using a stacking analysis. We then deduce the mean intensity of the radiation field and the mean dust mass in these objects using the \citet{Draine2007} model (Sect.\,\ref{stackfit}). We discuss the observed evolution of these quantities in Sect.\,\ref{results} and the consequences on the nature of star formation processes at high redshift in Sect.\,\ref{discussion}. Throughout this paper, we adopt a $\Lambda$CDM cosmology with $\Omega_m = 0.3$, $\Omega_\Lambda = 0.7$, $H_0 = 70$\,km/s/Mpc and a \citet{Chabrier2003} initial mass function (IMF).\\

\section{Data}

\label{data}

\subsection{Stellar mass and photometric redshift catalog using UltraVISTA data}

\label{masscat}

Deep Y, J, H, and K$_{\rm s}$ data (m$_{\rm AB, 5\sigma} \sim$ 25 for the Y band and 24 for the others) were produced by the UltraVISTA survey \citep{McCracken2012}. The photometric redshift and the stellar mass of the detected galaxies were estimated using Le PHARE \citep{Arnouts1999,Ilbert2006} as described in \citet{Ilbert2013}. The precision of the photometric redshifts at 1.5$<$z$<$4 is $\sigma_{\rm \Delta z / (1+z)}$ = 0.03. According to \citet{Ilbert2013}, this catalog is complete down to $10^{10.26}$\,M$_\odot$ at z$<$4. X-ray detected active galactic nuclei (AGNs) are also removed from our sample of star-forming galaxies, since the mid-infrared emission of these objects could be strongly affected the AGN.  Luminous X-ray obscured AGNs might still be present in the sample. However, their possible presence appear to have limited impact on our work as no mid-infrared excess is observed in the average SEDs measured by stacking (see Fig.\,\ref{fig:sedms} and \ref{fig:sedsb} and Sect.\,\ref{results}).\\

As this paper studies star-forming galaxies, we focused only on star-forming galaxies selected following the method of \citet{Ilbert2010} based on the rest-frame $\rm NUV-r^{+}$ versus $\rm r^{+}-J$ and similar to the UVJ criterion of \citep{Williams2009}. The flux densities in each rest-frame band are extrapolated from the closest observer-frame band to minimize potential biases induced by the choice of template library. At z$>$1.5, 40-60\,\% of the objects classified as passive by this color criterion have a sSFR$>10^{-11}$\,yr$^{-1}$ according to the SED fitting of the optical/near-IR data \citep[][their Fig.\,3]{Ilbert2013}. However, the sSFRs obtained by SED fitting are highly uncertain, because of the degeneracies with the dust attenuation. These peculiar objects are at least 10 times less numerous than the color-selected star-forming sample in all redshift bins. Including them or not in the sample has a negligible impact ($\sim0.25$\,$\sigma$) on the mean SEDs measured by stacking (see Sect.\,\ref{stackfit}). We thus based our study only on the color-selected population for simplicity.\\

\subsection{\textit{Spitzer}/MIPS data}

The COSMOS field (2 deg$^2$) was observed by \textit{Spitzer} at 24\,$\mu$m with the multiband imaging photometer (MIPS). A map and a catalog combined with the optical and near-IR data was produced from these observations \citep{Le_Floch2009}. The 1$\sigma$ point source sensitivity is $\sim$15\,$\mu$Jy and the full width at half maximum (FWHM) of the point spread function (PSF) is $\sim$6".\\

\subsection{\textit{Herschel}/PACS data}

The PACS (photodetecting array camera and spectrometer, \citealt{Poglitsch2010}) evolutionary probe survey (PEP, \citealt{Lutz2011}) mapped the COSMOS field with the \textit{Herschel}\footnote{{\it Herschel} is an ESA space observatory with science instruments provided by European-led Principal Investigator consortia and with important participation from NASA.} space observatory \citep{Pilbratt2010} at 100 and 160\,$\mu$m with a point-source sensitivity of 1.5 mJy and 3.3 mJy and a PSF FWHM of 7.7" and 12", respectively. Sources and fluxes of the PEP catalog were extracted using the position of 24\,$\mu$m sources as a prior. This catalog is used only to select strong starbursts up to z$\sim$3. The 24\,$\mu$m prior should not induce any incompleteness of the strong-starburst sample, since their minimum expected 24\,$\mu$m flux is at least 2 times larger than the detection limit at this wavelength \footnote{The minimum expected flux for our mass-selected sample of strong starbursts is computed using the three-dot-dash curve in Fig.\,\ref{SBcomp} and the \citet{Magdis2012b} starburst template.}.\\

\subsection{\textit{Herschel}/SPIRE data}

We also used \textit{Herschel} data at 250\,$\mu$m, 350\,$\mu$m, and 500\,$\mu$m taken by the spectral and photometric imaging receiver (SPIRE, \citealt{Griffin2010}) as part of the \textit{Herschel} multitiered extragalactic survey (HerMES, \citealt{Oliver2012}). The FWHM of the PSF is 18.2", 24.9", and 36.3", the 1\,$\sigma$ instrumental noise is 1.6, 1.3, and 1.9\,mJy/beam, and the 1\,$\sigma$ confusion noise is 5.8, 6.3, and 6.8\,mJy/beam \citep{Nguyen2010} at 250\,$\mu$m, 350\,$\mu$m, and 500\,$\mu$m, respectively. In this paper, we used the sources catalog extracted using as a prior the positions, the fluxes, the redshifts, and mean colors measured by stacking of 24\,$\mu$m sources, as described in \citet{Bethermin2012b}.\\

\subsection{LABOCA data}

The COSMOS field was mapped at 870\,$\mu$m by the large APEX bolometer Camera (LABOCA) mounted on the Atacama Pathfinder Experiment (APEX) telescope\footnote{APEX project IDs: 080.A--3056(A), 082.A--0815(A) and 086.A--0749(A).} (PI: Frank Bertoldi, Navarrete et al. in prep.). We retrieved the raw data from the ESO Science Archive facility and reduced them with the publicly available CRUSH (version 2.12--2) pipeline \citep{Kovacs2006,Kovacs2008}. We used the algorithm settings optimized for deep field observations\footnote{More details on the CRUSH settings can be found at: \url{http://www.submm.caltech.edu/$\sim$sharc/crush/v2/README}}. The output of CRUSH includes an intensity map and a noise map. The mapped area extends over approximately 1.4 square degrees with a non-uniform noise that increases toward the edges of the field. In this work we use the inner $\sim$0.7\,deg$^2$ of the map where a fairly uniform sensitivity of $\sim$4.3 mJy/beam is reached (Pannella et al. in prep.) with a smoothed beam size of $\sim$27.6". Contrary to SPIRE data, which are confusion limited, LABOCA data are noise limited and the maps are thus beam-smoothed to minimize their RMS.

\subsection{AzTEC data}

An area of 0.72\,deg$^2$ was scanned by the AzTEC bolometer camera mounted on the Atacama submillimeter telescope experiment (ASTE). The sensitivity in the center of the field is 1.23\,mJy RMS and the PSF FWHM after beam-smoothing is 34" \citep{Aretxaga2011}.\\

\section{Methods}

\label{stackfit}

\subsection{Sample selection}

\begin{figure}
\centering
\includegraphics{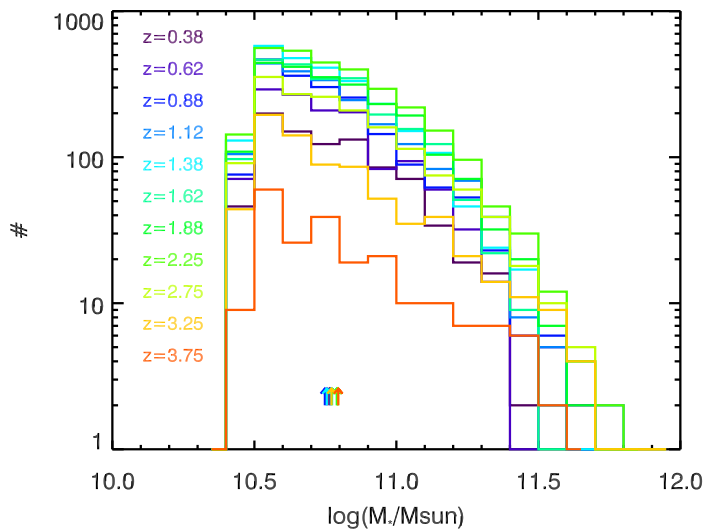}
\caption{\label{fig:massdistr} Stellar mass distribution of our samples of star-forming galaxies in the various redshift bins we used. Only galaxies more massive than our cut of $3\times 10^{10}\,\rm M_\odot$ are represented. The first bin contain fewer objects than the second one because our cut fall at the middle of the first one. The arrows indicate the mean stellar mass in each redshift bin.}
\end{figure}

In this paper, we base our analysis on mass-selected samples of star-forming galaxies (see Sect.\,\ref{masscat}). We chose the same stellar mass cut of $3 \times 10^{10}$\,M$_\odot$ at all redshifts to be complete up to z$\sim$4. We could have used a lower mass cut at lower redshifts, but we chose this single cut for all redshifts to be able to interpret the observed evolution of the various physical parameters of the galaxies in our sample in an easier way. This cut is slightly higher than the 90\% completeness limit at z$\sim$4 cited in \citet[][1.8$\times$10$^{10}$\,M$_\odot$]{Ilbert2013} and implies an high completeness of our sample, which limits potential biases induced by the input catalog on the results of our stacking analysis \citep[e.g.,][]{Heinis2013}. The exact choice of our stellar mass cut has negligible impact on the mean SEDs measured by stacking: we tested a mass cut of $2 \times 10^{10}$\,M$_\odot$ and $5 \times 10^{10}$\,M$_\odot$ and found that, after renormalization at the same L$_{\rm IR}$, the SEDs are similar ($\chi_{\rm red}^2$ = 0.57 and 0.79, respectively). These results agree with \citet{Magdis2012b}, who did not find any evidence of a dependence of the main-sequence SED on stellar mass at fixed redshift. The mass distribution of star-forming galaxies does not vary significantly with redshift, except in normalization (\citealt{Ilbert2013} and Fig.\,\ref{fig:massdistr}). The average stellar mass at all redshifts is between $10^{10.75}$\,M$_\odot$ and $10^{10.80}$\,M$_\odot$ (Fig.\,\ref{fig:massdistr} and Table\,\ref{tab:physpar}).\\

Star-forming galaxies whose stellar mass is larger than our cut do not correspond to the same populations at z=4 and z=0. The massive objects at z=4 are formed in dense environments, corresponding to the progenitors of today's clusters and massive groups \citep[e.g.,][]{Conroy2009,Moster2010,Behroozi2012a,Bethermin2013,Bethermin2014}. Most of these objects are in general quenched between z=4 and z=0 \citep[e.g.,][]{Peng2010}. In contrast, our mass cut at z=0 corresponds to Milky-Way-like galaxies. At all redshift, this cut is just below the mass corresponding to the maximal efficiency of star formation inside halos (defined here as the ratio between stellar mass and halo mass, \citealt{Moster2010,Behroozi2010,Bethermin2012b,Wang2013,Moster2013}).\\

Our stellar mass cut is slightly below the knee of the mass function of star-forming galaxies \citep{Ilbert2013}. The population we selected thus hosts the majority ($>$50\%) of the stellar mass in star-forming galaxies. Since there is a correlation between stellar mass and SFR, we are thus probing the population responsible for a large fraction the star formation (40-65\% depending on the redshift according to the \citealt{Bethermin2012b} model, see also \citealt{Karim2011}). Our approach is thus different from \citet{Santini2014} who explore in detail how the SEDs evolve at z$<$2.5 in the SFR-M$_\star$ plane using a combination of UV-derived and 24\,$\mu$m-derived SFRs. We aim to push our analysis to higher redshifts and we thus use this more simple and redshift-invariant selection to allow an easier interpretation and to limit potential selection biases. In addition to this mass selection, we divide our sample by intervals of redshift. The choice of their size is a compromise between large intervals to have a good signal-to-noise ratio at each wavelength and small intervals to limit the broadening of the SEDs because of redshift evolution within the broad redshift bin.\\

We also removed strong starbursts from our sample (sSFR$>$10 sSFR$_{\rm MS}$) and studied them separately. These objects are selected using the photometric catalogs described in Sect.\,\ref{data}. For the sources which are detected at 5\,$\sigma$ at least in two \textit{Herschel} bands, we fitted the SEDs with the template library of \citet{Magdis2012b} allowing the mean intensity of the radiation field $\langle U \rangle$ to vary by $\pm$0.6\,dex (3$\sigma$ of the scatter used in the \citealt{Bethermin2012c} model). These criteria of two detections at different wavelengths and the high reliability of the detections prevent biasing of the starbursts towards positive fluctuations of the noise in the maps and limit the flux boosting effect. We then estimated the SFR from the infrared luminosity, L$_{\rm IR}$, using the \citet{Kennicutt1998} relation. We performed a first analysis using the same evolution of the main-sequence (sSFR$_{\rm MS}$ versus z) as in \citet{Bethermin2012c} to select sSFR$>$10 sSFR$_{\rm MS}$ objects. We then fit the measured evolution of the main-sequence found by a first stacking analysis (see Sect.\,\ref{sect:stacking} and Sect.\,\ref{sect:sedfit}) to prepare the final sample for our analysis. We could have chosen a lower sSFR cut corresponding to 4 times the value at the center of the main-sequence as in \citet{Rodighiero2011}, but the sample would be incomplete at z$>$1 because of the flux limit of the infrared catalogs.\\

\begin{figure}
\centering
\includegraphics{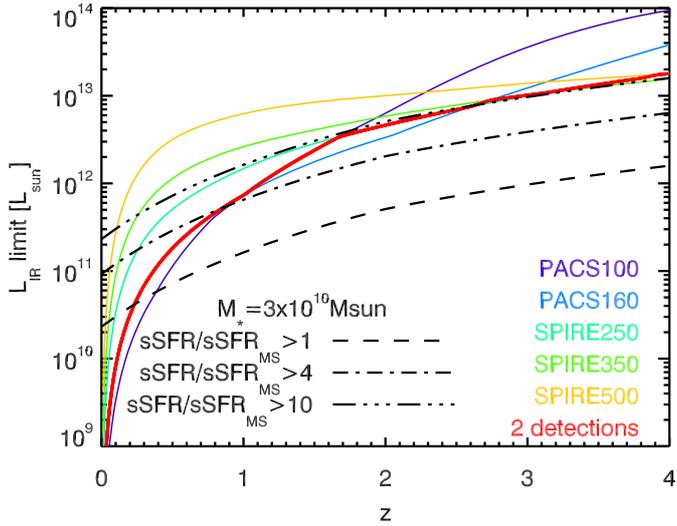}
\caption{\label{SBcomp} The thick red solid line represents the luminosity limit corresponding to a criterion of a 5\,$\sigma$ detection in at least two \textit{Herschel} bands. The other solid lines are the limits for a detection at only one given wavelength (purple for 100\,$\mu$m, blue for 160\,$\mu$m, turquoise for 250\,$\mu$m, green for 350\,$\mu$m, orange for 500\,$\mu$m). The dashed, dot-dash, and three-dot-dash lines indicate the infrared luminosity of a galaxie of $3 \times 10^{10}$\,M$_\odot$ (our mass cut) at the center of the main sequence, a factor of 4 above it, and a factor of 10 above it, respectively.}
\end{figure}

Fig.\,\ref{SBcomp} shows the luminosity limit corresponding to a detection at 5\,$\sigma$ at two wavelengths or more. This was computed using both the starburst and the main-sequence templates of the \citet{Magdis2012b} SED library. This library contains different templates for main-sequence and starburst galaxies. The main-sequence template evolves with redshift, but not the starburst one. The lines correspond to the highest luminosity limit found using these two templates for each wavelength, which is the most pessimistic case. We also computed the infrared luminosity associated with a galaxy of $3\times 10^{10}\,\rm M_\odot$, i.e., our mass limit, on the main sequence (dashed line), a factor of 4 above it (dot-dash line), and a factor of 10 above it (three-dot-dash line). All the M$_\star > 3\times 10^{10}\,\rm M_\odot$ strong starbursts (sSFR$>$10 sSFR$_{\rm MS}$) should thus be detected in two or more \textit{Herschel} bands below z=4. There is only one starburst detected in the 3$<$z$<$4 bin. We thus do not analyze this bin, because of its lack of statistical significance. The other bins contain 3, 6, 6, and 8 strong starbursts, respectively, by increasing redshift.\\

The sample of main-sequence galaxies is contaminated by the starbursts which have sSFR$<$10 sSFR$_{\rm MS}$ . We expect that this contamination is negligible, since the contribution of all starbursts to the infrared luminosity density is lower than 15\% \citep{Rodighiero2011,Sargent2012}. To check this hypothesis, we statistically corrected for the contribution of the remaining starbursts with sSFR$<$10 sSFR$_{\rm MS}$ based on the \citet{Bethermin2012b} counts model. We assumed both the SED library used for the model and the average SED of strong starbursts found in this study. We found that this statistical subtraction only affected our measurements at most at the 0.2$\sigma$ level. Consequently, we have neglected this contamination in the rest of our study.\\

\begin{table*}
\centering
\caption{\label{tab:fluxes} Summary of our flux density measurements by stacking}
\begin{tabular}{ccccccccc}
\hline
\hline
Redshift & S$_{24}$ & S$_{100}$ & S$_{160}$ & S$_{250}$ & S$_{350}$ & S$_{500}$ & S$_{850}$ & S$_{1100}$\\
\hline
 & $\mu$Jy & mJy& mJy& mJy& mJy& mJy& mJy & mJy \\
\hline
\multicolumn{9}{c}{Main-sequence sample} \\
\hline
0.25$<$z$<$0.50 & 410$\pm$23 & 11.87$\pm$0.76 & 23.30$\pm$1.49 & 12.54$\pm$0.97 & 6.43$\pm$0.53 & 2.64$\pm$0.32 & -0.18$\pm$0.23 & 0.21$\pm$0.08 \\
0.50$<$z$<$0.75 & 247$\pm$13 & 6.37$\pm$0.43 & 13.82$\pm$0.86 & 9.45$\pm$0.72 & 5.88$\pm$0.46 & 2.57$\pm$0.25 & 0.54$\pm$0.15 & 0.18$\pm$0.06 \\
0.75$<$z$<$1.00 & 221$\pm$10 & 4.19$\pm$0.26 & 9.79$\pm$0.60 & 7.75$\pm$0.59 & 5.92$\pm$0.45 & 3.06$\pm$0.25 & 0.53$\pm$0.19 & 0.30$\pm$0.06 \\
1.00$<$z$<$1.25 & 144$\pm$7 & 3.31$\pm$0.23 & 8.22$\pm$0.50 & 6.93$\pm$0.53 & 5.78$\pm$0.46 & 3.00$\pm$0.25 & 0.21$\pm$0.15 & 0.30$\pm$0.05 \\
1.25$<$z$<$1.50 & 96$\pm$5 & 2.36$\pm$0.14 & 6.70$\pm$0.42 & 5.99$\pm$0.45 & 5.46$\pm$0.41 & 3.17$\pm$0.25 & 0.44$\pm$0.13 & 0.32$\pm$0.04 \\
1.50$<$z$<$1.75 & 110$\pm$6 & 1.80$\pm$0.12 & 4.81$\pm$0.33 & 4.79$\pm$0.38 & 4.64$\pm$0.36 & 3.00$\pm$0.25 & 0.54$\pm$0.11 & 0.34$\pm$0.04 \\
1.75$<$z$<$2.00 & 113$\pm$5 & 1.31$\pm$0.10 & 3.51$\pm$0.25 & 4.10$\pm$0.32 & 4.11$\pm$0.33 & 2.94$\pm$0.24 & 0.72$\pm$0.12 & 0.32$\pm$0.04 \\
2.00$<$z$<$2.50 & 101$\pm$5 & 1.16$\pm$0.08 & 3.28$\pm$0.22 & 4.17$\pm$0.32 & 4.38$\pm$0.34 & 3.25$\pm$0.25 & 0.73$\pm$0.12 & 0.48$\pm$0.04 \\
2.50$<$z$<$3.00 & 59$\pm$3 & 0.79$\pm$0.07 & 2.59$\pm$0.22 & 3.41$\pm$0.29 & 3.85$\pm$0.31 & 3.03$\pm$0.26 & 0.87$\pm$0.17 & 0.55$\pm$0.05 \\
3.00$<$z$<$3.50 & 47$\pm$5 & 0.61$\pm$0.10 & 2.28$\pm$0.33 & 2.90$\pm$0.30 & 3.65$\pm$0.35 & 2.95$\pm$0.31 & 0.56$\pm$0.18 & 0.44$\pm$0.07 \\
3.50$<$z$<$4.00 & 29$\pm$7 & 0.22$\pm$0.20 & 1.68$\pm$0.55 & 2.60$\pm$0.45 & 3.01$\pm$0.51 & 2.52$\pm$0.50 & 0.24$\pm$0.33 & 0.30$\pm$0.14 \\
\hline
\multicolumn{9}{c}{Strong-starburst sample} \\
\hline
0.50$<$z$<$1.00 & 1241$\pm$329 & 57.48$\pm$15.98 & 86.33$\pm$18.31 & 41.57$\pm$7.83 & 16.52$\pm$3.53 & 9.64$\pm$4.73 & 6.91$\pm$5.92 & 2.40$\pm$1.57 \\
1.00$<$z$<$1.50 & 264$\pm$77 & 30.59$\pm$3.26 & 64.44$\pm$6.97 & 38.44$\pm$4.92 & 24.79$\pm$3.98 & 13.90$\pm$4.97 & 0.12$\pm$2.62 & 1.36$\pm$0.78 \\
1.50$<$z$<$2.00 & 912$\pm$179 & 23.51$\pm$5.04 & 62.46$\pm$13.80 & 42.47$\pm$8.02 & 30.99$\pm$9.27 & 21.46$\pm$7.09 & 2.10$\pm$3.37 & 3.90$\pm$1.16 \\
2.00$<$z$<$3.00 & 629$\pm$193 & 13.15$\pm$4.91 & 39.56$\pm$7.77 & 32.25$\pm$4.37 & 35.72$\pm$5.40 & 28.52$\pm$5.20 & 7.98$\pm$2.97 & 5.08$\pm$1.02 \\
\hline
\end{tabular}
\end{table*}

\begin{table*}
\centering
\caption{\label{tab:physpar} Summary of the average physical parameters of our samples}
\begin{tabular}{cccccccc}
\hline
\hline
Redshift & log(M$_\star$) & log(L$_{\rm IR}$) & SFR & log(M$_{\rm dust}$) & $\langle U \rangle$ & log(M$_{\rm mol}$) &  f$_{\rm mol}$ \\
\hline
 & log(M$_\odot$) & log(L$_\odot$) & M$_\odot$/yr & log(M$_\odot$) & & log(M$_\odot$) & \\
\hline
\multicolumn{8}{c}{Main-sequence sample} \\
\hline
\vspace{0.1cm}
0.25$<$z$<$0.50 & 10.77 & 10.92$_{-0.04}^{+0.03}$ & 8.3$_{-0.7}^{+0.6}$ & 8.09$_{-0.16}^{+0.12}$ & 5.50$_{-1.50}^{+3.10}$ & 10.04$_{-0.22}^{+0.19}$ & 0.16$_{-0.06}^{+0.07}$ \\
0.50$<$z$<$0.75 & 10.76 & 11.19$_{-0.04}^{+0.08}$ & 15.6$_{-1.5}^{+3.3}$ & 8.24$_{-0.15}^{+0.19}$ & 7.23$_{-2.47}^{+3.82}$ & 10.23$_{-0.21}^{+0.24}$ & 0.23$_{-0.07}^{+0.11}$ \\
0.75$<$z$<$1.00 & 10.75 & 11.45$_{-0.09}^{+0.07}$ & 27.9$_{-5.4}^{+4.7}$ & 8.44$_{-0.24}^{+0.16}$ & 7.80$_{-2.69}^{+5.44}$ & 10.48$_{-0.28}^{+0.22}$ & 0.35$_{-0.13}^{+0.12}$ \\
1.00$<$z$<$1.25 & 10.77 & 11.56$_{-0.04}^{+0.10}$ & 36.4$_{-3.3}^{+9.7}$ & 8.29$_{-0.11}^{+0.28}$ & 15.05$_{-6.68}^{+5.74}$ & 10.34$_{-0.18}^{+0.32}$ & 0.27$_{-0.07}^{+0.16}$ \\
1.25$<$z$<$1.50 & 10.76 & 11.69$_{-0.04}^{+0.07}$ & 48.6$_{-4.2}^{+8.9}$ & 8.37$_{-0.10}^{+0.22}$ & 16.52$_{-6.47}^{+5.45}$ & 10.46$_{-0.18}^{+0.26}$ & 0.33$_{-0.08}^{+0.15}$ \\
1.50$<$z$<$1.75 & 10.77 & 11.77$_{-0.05}^{+0.05}$ & 58.9$_{-5.9}^{+7.5}$ & 8.45$_{-0.21}^{+0.18}$ & 16.96$_{-6.15}^{+10.90}$ & 10.55$_{-0.26}^{+0.23}$ & 0.37$_{-0.13}^{+0.13}$ \\
1.75$<$z$<$2.00 & 10.79 & 11.81$_{-0.03}^{+0.05}$ & 64.4$_{-4.3}^{+8.2}$ & 8.49$_{-0.25}^{+0.18}$ & 16.96$_{-6.15}^{+15.24}$ & 10.63$_{-0.29}^{+0.23}$ & 0.41$_{-0.15}^{+0.13}$ \\
2.00$<$z$<$2.50 & 10.79 & 11.99$_{-0.02}^{+0.03}$ & 97.4$_{-5.3}^{+7.7}$ & 8.53$_{-0.19}^{+0.13}$ & 22.58$_{-6.27}^{+14.42}$ & 10.81$_{-0.24}^{+0.20}$ & 0.51$_{-0.13}^{+0.11}$ \\
2.50$<$z$<$3.00 & 10.80 & 12.11$_{-0.04}^{+0.03}$ & 130.0$_{-12.6}^{+10.7}$ & 8.48$_{-0.11}^{+0.23}$ & 33.75$_{-14.29}^{+12.85}$ & 10.88$_{-0.18}^{+0.27}$ & 0.55$_{-0.10}^{+0.15}$ \\
3.00$<$z$<$3.50 & 10.77 & 12.25$_{-0.05}^{+0.05}$ & 178.5$_{-18.4}^{+22.4}$ & 8.48$_{-0.12}^{+0.10}$ & 48.99$_{-11.32}^{+23.99}$ & 10.99$_{-0.19}^{+0.18}$ & 0.62$_{-0.11}^{+0.09}$ \\
3.50$<$z$<$4.00 & 10.80 & 12.34$_{-0.12}^{+0.07}$ & 219.0$_{-54.4}^{+40.2}$ & 8.39$_{-0.50}^{+0.33}$ & 72.98$_{-36.98}^{+167.95}$ & 11.06$_{-0.52}^{+0.36}$ & 0.65$_{-0.29}^{+0.16}$ \\
\hline
\multicolumn{8}{c}{Strong-starburst sample} \\
\hline
\vspace{0.1cm}
0.50$<$z$<$1.00 & 10.57 & 12.25$_{-0.08}^{+0.08}$ & 179.1$_{-150.5}^{+215.0}$ & 8.65$_{-0.04}^{+0.19}$ & 29.80$_{-11.77}^{+9.60}$ & 10.04$_{-0.24}^{+0.30}$ & 0.29$_{-0.10}^{+0.16}$ \\
1.00$<$z$<$1.50 & 10.60 & 12.55$_{-0.05}^{+0.03}$ & 350.8$_{-314.5}^{+376.4}$ & 8.99$_{-0.01}^{+0.09}$ & 26.92$_{-6.92}^{+2.88}$ & 10.23$_{-0.23}^{+0.25}$ & 0.45$_{-0.13}^{+0.14}$ \\
1.50$<$z$<$2.00 & 10.64 & 12.93$_{-0.18}^{+0.07}$ & 860.1$_{-567.4}^{+1006.8}$ & 9.24$_{-0.09}^{+0.62}$ & 37.68$_{-28.40}^{+11.32}$ & 10.48$_{-0.25}^{+0.66}$ & 0.58$_{-0.14}^{+0.28}$ \\
2.00$<$z$<$3.00 & 10.69 & 13.10$_{-0.24}^{+0.07}$ & 1260.0$_{-728.1}^{+1487.1}$ & 9.64$_{-0.47}^{+0.37}$ & 22.22$_{-12.94}^{+50.77}$ & 10.34$_{-0.52}^{+0.44}$ & 0.75$_{-0.28}^{+0.14}$ \\
\hline
\end{tabular}
\end{table*}

\begin{figure*}
\centering
\includegraphics{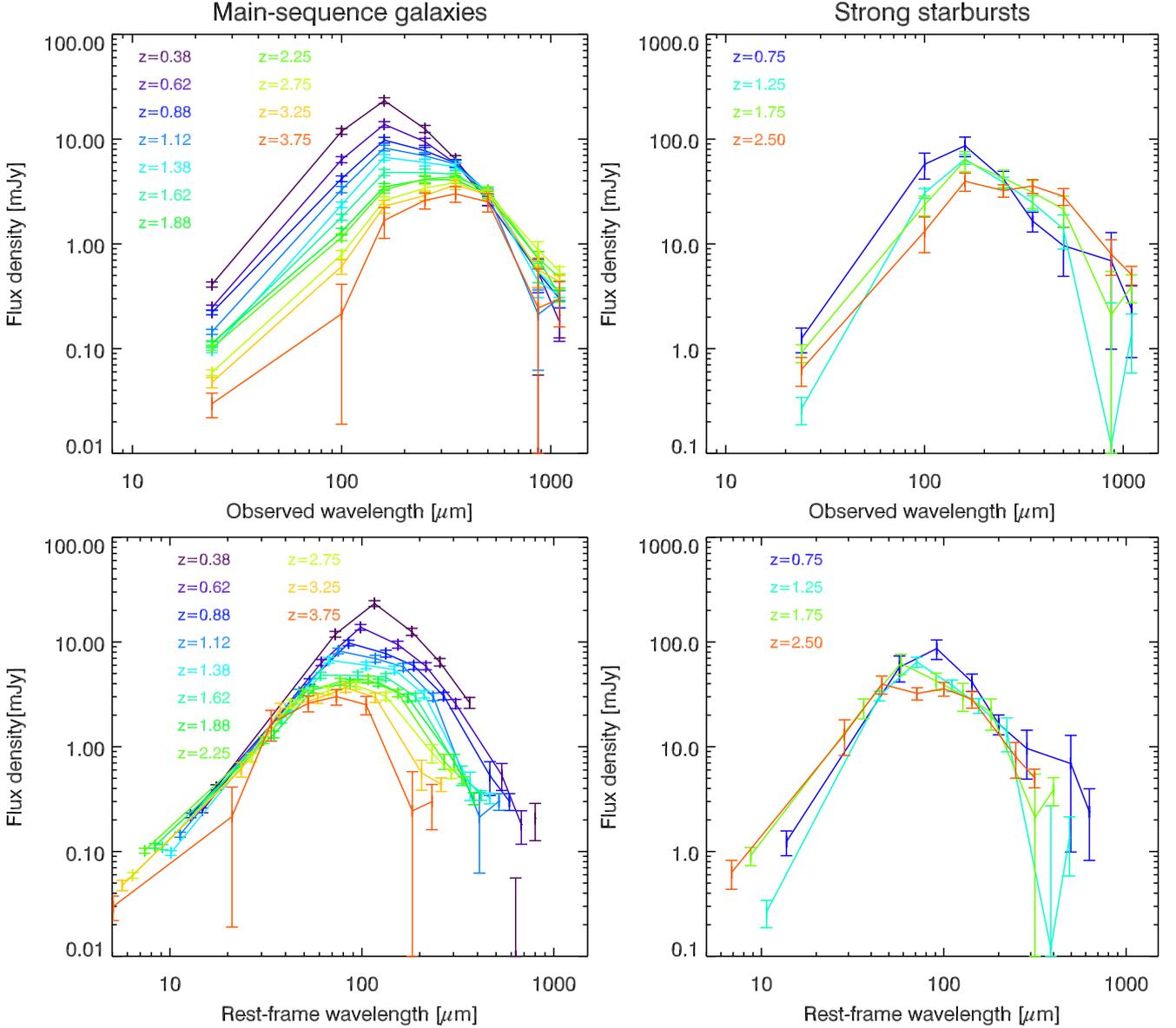}
\caption{\label{fig:sedobs} Mean flux density as a function of wavelength (observed wavelength in the top panels and rest-frame wavelength in the bottom panels) at various redshifts (see color coding). The left panels show the mean SEDs of the main-sequence sample and the right panels those of the strong starbursts.}
\end{figure*}

\begin{figure*}
\centering
\includegraphics{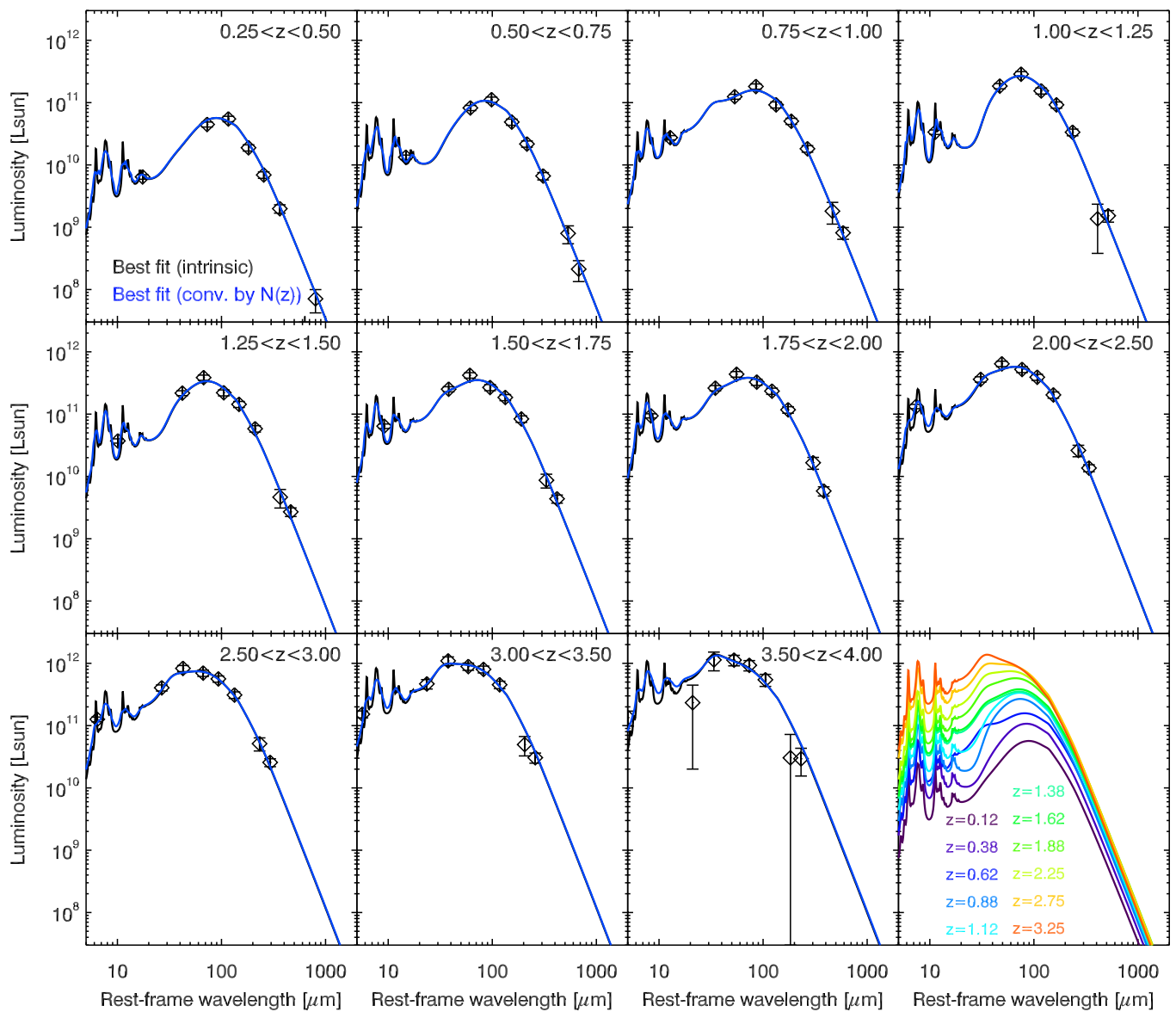}
\caption{\label{fig:sedms} Rest-frame mean spectral energy distribution of our selection of massive, star-forming galaxies at various redshift measured by stacking analysis. The data points are fitted using the \citet{Draine2007} model. This model is convolved with the redshift distribution of the sources before being compared to the data. The black and blue lines represent the intrinsic and convolved SEDs, respectively. The bottom right corner summarizes the redshift evolution seen in our data.}
\end{figure*}

\begin{figure*}
\centering
\includegraphics{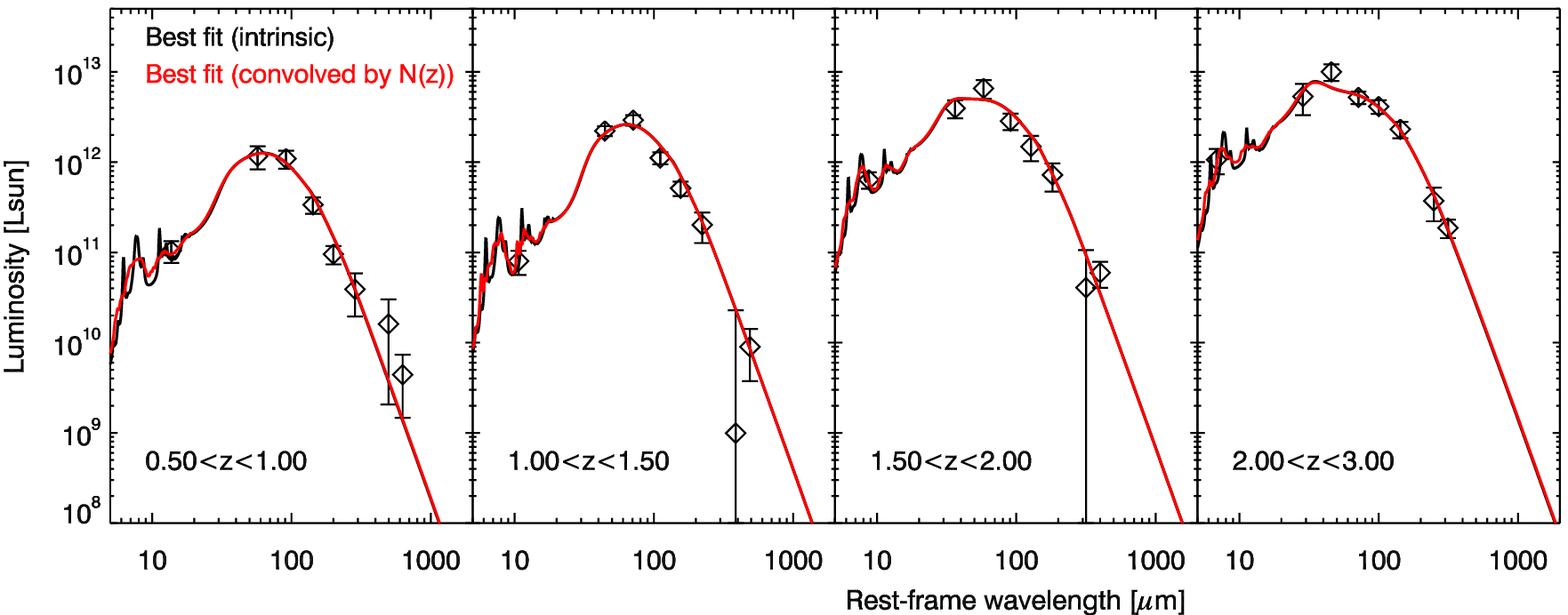}
\caption{\label{fig:sedsb} Rest-frame mean spectral energy distribution of our selection of strong starbursts at various redshift measured by stacking analysis. The data points are fitted using the \citet{Draine2007} model. This model is convolved with the redshift distribution of the sources before being compared to the data. The black and red lines show the intrinsic and convolved SEDs, respectively.}
\end{figure*}

\subsection{Stacking analysis}

\label{sect:stacking}

We use a similar stacking approach as in \citet{Magdis2012b} to measure the mean SEDs of our sub-samples of star-forming galaxies from the mid-infrared to the millimeter domain. Different methods are used at the various wavelength to optimally extract the information depending if the data are confusion or noise limited. At 24\,$\mu$m, 100\,$\mu$m, and 160\,$\mu$m, we produced stacked images using the IAS stacking library \citep{Bavouzet2008,Bethermin2010a}. The flux is then measured using aperture photometry with the same parameters and aperture corrections as \citet{Bethermin2010a} at 24\,$\mu$m. At 100\,$\mu$m and 160\,$\mu$m, we used a PSF fitting technique. A correction of 10\% is applied to take into account the effect of the filtering of the data on the photometric measurements of faint, non-masked sources \citep{Popesso2012}. At 250\,$\mu$m, 350\,$\mu$m, and 500\,$\mu$m, the photometric uncertainties are not dominated by instrumental noise but by the confusion noise caused by neighboring sources \citep{Dole2003,Nguyen2010}. We thus measured the mean flux of the sources computing the mean flux in the pixels centered on a stacked source following \citet{Bethermin2012b}. This method minimizes the uncertainties and a potential contamination caused by the clustering of galaxies \citep{Bethermin2010b}. Finally, we used the same method, but on the beam-convolved map, for LABOCA and AzTEC data as they are noise limited and lower uncertainties are obtained after this beam smoothing. LABOCA and AzTEC maps do not cover the whole area. We thus only stack sources in the covered region to compute the mean flux densities of our various sub-samples. The source selection criteria being exactly the same inside and outside the covered area, this should not introduce any bias.\\

These stacking methods can be biased if the stacked sources are strongly clustered or very faint. This bias is caused by the greater probability of finding a source close to another one in the stacked sample compared to a random position. This effect has been discussed in detail by several authors \citep[e.g.,][]{Bavouzet2008,Bethermin2010b,Kurczynski2010,Bethermin2012b,Bourne2012,Viero2013b}. In \citet{Magdis2012b}, the authors estimated that this bias is lower than the 1$\sigma$ statistical uncertainties and was not corrected. The number of sources to stack in COSMOS compared to the GOODS fields used by \citet{Magdis2012b} is much larger and hence the signal-to-noise ratio is much better. The bias caused by clustering is thus non-negligible in COSMOS. Because of the complex edge effects caused by the absence of data around bright stars, the methods using the position of the sources to deblend the contamination caused by the clustering cannot be applied \citep{Kurczynski2010,Viero2013b}. Consequently, we developed a method based on realistic simulations of the \textit{Spitzer}, \textit{Herschel}, LABOCA, and AzTEC maps to correct this effect, which induces biases up to 50\% at 500\,$\mu$m around z$\sim$2. The technical details and discussion of these corrections are presented in Appendix\,\ref{Annexestacking}.\\

The uncertainties on the fluxes are measured using a bootstrap technique \citep{Jauzac2011}. This method takes into account both the errors coming from the instrumental noise, the confusion, and the sample variance of the galaxy population \citep{Bethermin2012b}. These uncertainties are combined quadratically with those associated with the calibration and the clustering correction.\\

\subsection{Mean physical properties from SED fitting}

\label{sect:sedfit}

We interpreted our measurements of the mean SEDs using the \citet{Draine2007} model as in \citet{Magdis2012b}. This model, developed initially to study the interstellar medium in the Milky Way and in nearby galaxies, takes into account the heterogeneity of the intensity of the radiation field. The redshift slices we used have a non-negligible width. To account for this, we convolve the model by the redshift distribution of the galaxies before fitting the data. The majority of the redshifts in our sample are photometric. We thus sum the probability distribution function (PDF) of the redshifts of all the sources in a sub-sample to estimate its intrinsic redshift distribution. The uncertainties on the physical parameters are estimated using the same Monte Carlo method as in \citet{Magdis2012b}. The uncertainties on each parameter takes into account the potential degeneracies with the others, i.e., they are the marginalized uncertainties on each individual parameters. Our good sampling of the dust SEDs (8 photometric points between 24\,$\mu$m and 1.1\,mm including at least six detections) allows us to break the degeneracy between the dust temperature and the dust mass which is found if only (sub-)mm datapoints are used.\\ 

Instead of using the three parameters describing the distribution of the intensity of the radiation field U of the \citet{Draine2007} model (the minimal radiation field $\rm U_{min}$, the maximal one $\rm U_{max}$, and the slope of the assumed power-law distribution between these two values $\alpha$), we considered only the mean intensity of the radiation field $\langle U \rangle$ for simplicity. The other parameters derived from the fit and used in this paper are the bolometric infrared luminosity integrated between 8 and 1000\,$\mu$m (L$_{\rm IR}$)  and the dust mass (M$_d$). The SFR is derived from L$_{\rm IR}$ using the \citet{Kennicutt1998} conversion factor ($1 \times 10^{-10}$\,$\rm M_\odot \, yr^{-1} \, L_\odot^{-1}$ after conversion from Salpeter to Chabrier IMF), since the dust-obscured star formation vastly dominates the unobscured component at z$<$4 given the mass-scale considered \citep{Heinis2013,Heinis2014,Pannella2014}. The sSFR is computed using the later SFR and the mean stellar mass extracted from the \citet{Ilbert2013} catalog. The uncertainties on the derived physical parameters presented in the various figures and tables of this paper are the uncertainties on the average values. The dispersion of physical properties inside a population is difficult to measure by stacking and we did not try to compute it in this paper (see Sect.\,\ref{discussion}).\\

The residuals of these fits are presented in Appendix\,\ref{sect:residuals}. Tables\,\ref{tab:fluxes} and \ref{tab:physpar} summarize the average photometric measurements and the recovered physical parameters, respectively.\\

\section{Results}

\label{results}

\subsection{Evolution of the mean SED of star-forming galaxies}

Figure\,\ref{fig:sedobs} summarizes the results of our stacking analysis. For the main-sequence sample, the flux density varies rapidly with redshift in the PACS 100\,$\mu$m band, while it is almost constant in the SPIRE 500$\mu$m band. The peak of the flux density distribution in the rest frame moves from $\sim$120$\mu$m to 70$\mu$m between z=0 and z=4. This shift with redshift was already observed at z$\lesssim$2 for mass-selected stacked samples \citep{Magdis2012b} or a \textit{Herschel}-detected sample \citep{Lee2013,Symeonidis2013}. We found no evidence of an evolution of the position of this peak ($\sim$70$\mu$m) for the sample of strong starbursts.\\

Figure\,\ref{fig:sedms} and \ref{fig:sedsb} show the mean intrinsic luminosity (in $\nu$L$_\nu$ units, the peak of the SEDs is thus shifted toward shorter wavelengths compared with L$_\nu$ units) of our samples of massive star-forming galaxies (since this sample is dominated by main-sequence galaxies, hereafter we call it main-sequence sample) and the fit by the \citet{Draine2007} model. We also observe a strong evolution of the position of the peak of the thermal emission of dust in main-sequence galaxies from $\sim$80\,$\mu$m at z$\sim$0.4 to $\sim$30\,$\mu$m at z$\sim$3.75 in $\nu$L$_\nu$ units. The SEDs of strong starbursts have a much more modest evolution (from 50\,$\mu$m at to 30\,$\mu$m). The mean luminosity of the galaxies also increases very rapidly with redshift for both main-sequence and strong starburst galaxies.\\

At z$>$2, we find that the peak of the dust emission tends to be broader than at lower redshift. The broadening of the mean SEDs induced by the size of the redshift bins has a major impact only on the mid-infrared, where the polycyclic aromatic hydrocarbon (PAH) features are washed out (see black and blue lines in Fig.\,\ref{fig:sedms} and \ref{fig:sedsb}), and cannot fully explain why the far-IR peak is broader at higher redshifts. The \citet{Draine2007} model reproduces this broadening by means of a higher $\gamma$ coefficient, i.e., a stronger contribution of regions with a strong heating of the dust. This is consistent with the presence of giant star-forming clumps in high-redshift galaxies \citep[e.g.,][]{Bournaud2007,Genzel2006}. The best-fit models at high z presents two breaks around 30\,$\mu$m and 150\,$\mu$m, which could be artefacts caused by the sharp cuts of the U distribution at its extremal values in the \citet{Draine2007} model.\\

\subsection{Evolution of the specific star formation rate}

\begin{figure}
\centering
\includegraphics{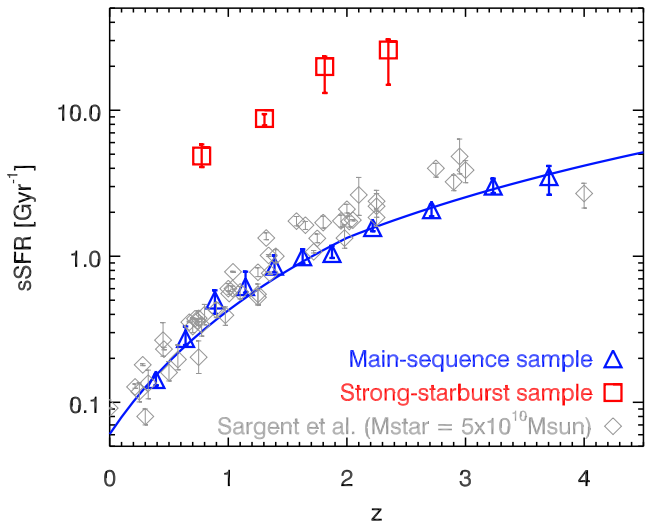}
\caption{\label{fig:ssfr} Evolution of the mean sSFR in main-sequence galaxies (blue triangles) and strong starbursts (red squares). The gray diamonds are a compilation of measurements at the same mass performed by \citet{Sargent2014}. The blue line is the best fit to our data.}
\end{figure}

From the fit of the SEDs, we can easily derive the evolution of the mean specific star formation rate of our mass-selected sample with redshift. The results are presented in Fig.\,\ref{fig:ssfr}. The strong starbursts lie about a factor of 10 above the main-sequence, demonstrating that this population is dominated by objects just above our cut of 10 sSFR$_{\rm MS}$. Our results can be fitted by an evolution in redshift as (0.061$\pm$0.006\,Gyr$^{-1}$)$\times$(1+z)$^{2.82\pm0.12}$ at z$<$2 and as (1+z)$^{2.2\pm0.3}$ at z$>$2. We compared our results with the compilation of measurements of \citet{Sargent2014} at M$_\star = 5 \times 10^{10}$\,M$_\odot$. At z$<$1.5, our results agree well with the previous measurements. Between z=1.5 and z=3.5, our new measurements follow the lower envelop of the previous measurements. This mild disagreement could have several causes.\\

First of all, the clustering effect was not taken into account by the previous analyses based on stacking. This effect is stronger at high redshift, because the bias\footnote{The bias $b$ is defined by $w_{\rm gal} = b^2 w_{\rm DM}$, where $w_{\rm gal}$ and $w_{\rm DM}$ are the projected two-point correlation function of galaxies and dark matter, respectively. The higher the bias is, the stronger is the clustering density of galaxies compared to dark matter.} of both infrared and mass-selected galaxies increases with redshift \citep[e.g.,][]{Bethermin2013}. In addition, the SEDs peak at a longer wavelength, where the bias is stronger owing to beam size (see Sect.\,\ref{sect:simu}). The tension with the results based on UV-detected galaxies could be explained by a slight incompleteness of the UV-detected samples at low sSFR or a small overestimate of the dust corrections. There could also be effects caused by the different techniques and assumptions used to determine the stellar masses in the various fields (star formation histories, PSF-homogenized photometry or not, presence of nebular emission in the highest redshift bins, template libraries, etc.). Finally, this difference could also be an effect of the variance. These discrepancies on the estimates of sSFRs will be discussed in detail in \citet{Schreiber2014}.\\

\label{fig:sSFR}

\subsection{Evolution of the mean intensity of the radiation field}

\label{sect:U}

\begin{figure}
\centering
\includegraphics{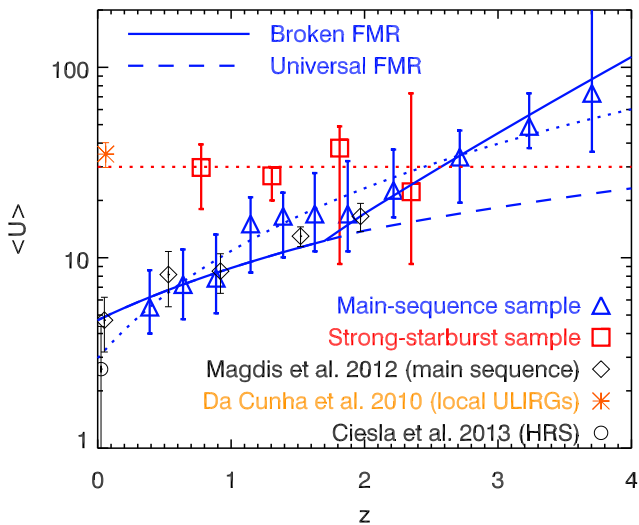}
\caption{\label{fig:U} Evolution of the mean intensity of the radiation field $\langle U \rangle$ in main-sequence galaxies (blue triangles) and strong starbursts (red squares). The black diamonds are the measurements presented in \citet{Magdis2012b} based on a similar analysis but in the GOODS fields. The orange asterisk is the mean value found for the local ULIRG sample of \citet{Da_Cunha2008b} (see also \citealt{Magdis2012b}). The black circle is the average value in HRS galaxies \citep{Ciesla2014}. The solid and dashed lines represent the evolutionary trends expected for a broken and universal FMR, respectively (see Sect.\,\ref{sect:U}). The blue dotted line is the best fit of the evolution of the main-sequence galaxies ($(3.0\pm1.1) \times (1+z)^{1.8 \pm 0.4}$) and the red dotter line the best fit of the strong starburst data by a constant ($31\pm3$).}
\end{figure}

The evolution of the mean intensity of the radiation field has different trends in main-sequence galaxies than in strong starbursts (see Fig.\,\ref{fig:U}). This quantity is strongly correlated to the temperature of the dust. We found a rising $\langle U \rangle$ with increasing redshift in main-sequence galaxies up to z=4 ($(3.0\pm1.1) \times (1+z)^{1.8\pm0.4}$), confirming and extending the finding of \citet{Magdis2012b} at higher redshift. Other studies \citep[e.g.,][]{Magnelli2013,Genzel2014} found an increase of the dust temperature with redshift in mass-selected samples.\\

The evolution of $\langle U \rangle$ we found can be understood from a few simple assumptions on the evolution of the gas metallicity and the star-formation efficiency (SFE) of galaxies. As shown by \citet{Magdis2012b}, $\langle U \rangle$ is proportional to L$_{\rm IR}$/M$_{\rm dust}$. We can also assume that 
\begin{equation}
L_{\rm IR} \propto \textrm{SFR} \propto M_{\rm mol}^{1/s},
\end{equation}
where the left-side of the proportionality is the well-established \citet{Kennicutt1998} relation. The right-side of the proportionality is the integrated version of the Schmidt-Kennicutt relation which links the SFR to the mass of molecular gas in a galaxy (M$_{\rm mol}$). \citet{Sargent2014} found a best-fit value for $s$ of 0.83 compiling a large set of public data about low- and high-redshift main-sequence galaxies. The molecular gas mass can also be connected to the gas metallicity Z and the dust mass \citep[e.g.,][]{Leroy2011,Magdis2012b},
\begin{equation}
M_{\rm dust} \propto Z(M_\star, \textrm{SFR}) \times M_{\rm mol},
\end{equation} 
where $Z(M_\star, \textrm{SFR})$ is the gas metallicity which can be connected to M$_\star$ and SFR through the fundamental metallicity relation (FMR, \citealt{Mannucci2010}). There is recent evidence showing that this relation breaks down at high redshifts. For instance, \citet{Troncoso2014} measured a $\sim$0.5\,dex lower normalization at z$\sim$3.4 compared to the functional form of the FMR at low redshift. \citet{Amorin2014} found the same offset in a lensed galaxy at z = 3.417. At z$\sim$2.3, \citet[][see also \citealt{Cullen2014}]{Steidel2014} found an offset of 0.34--0.38\,dex in the mass-metallicity relation and only half of this difference can be explained by the increase of SFR at fixed stellar mass using the FMR. Finally, a break in the metallicity relation is also observed in low mass (log(M$_\star$/M$_\odot$)$\sim$8.5) damped Lyman $\alpha$ absorbers around z=2.6 \citep{Moller2013}. In our computations, we consider two different relations: a universal FMR where metallicity depends only on M$_\star$ and SFR, and a FMR relation with a correction of $0.30 \times (1.7-z)$\,dex at z$>$1.7 (hereafter broken FMR), which agrees with the measurements cited previously. Combining these expressions,  we can obtain the following evolution:
\begin{equation}
\langle U \rangle \propto \frac{L_{\rm IR}}{M_{\rm dust}} \propto \frac{M_{\rm mol}^{\frac{1}{s}-1}}{Z(M_\star, \textrm{SFR})} \propto \frac{\textrm{SFR}^{1-s}}{Z(M_\star, \textrm{SFR})}.\\
\end{equation}
We computed the expected evolution of $\langle U \rangle$ using the fit to the evolution of sSFR presented in Sect.\,\ref{fig:sSFR} and assuming the mean stellar mass of our sample is $6\times10^{10}$\,M$_\odot$, the average mass of the main-sequence sample\footnote{We could have used the mean stellar masses in each redshift bin provided in Table\,\ref{tab:physpar}. However, assuming a single value of the stellar mass at all redshift has a negligible impact on the results and the tracks are smoother.}. We used the value of \citet{Magdis2012b} at z=0 to normalize our model. The results are presented in Fig.\,\ref{fig:U} for a universal and a broken FMR. The broken FMR is compatible with all of our data points at 1$\sigma$. The universal FMR implies a significant underestimation of $\langle U \rangle$ at high redshifts (3 and 2\,$\sigma$ in the two highest redshift bins).\\

We checked that the dust heating by the cosmic microwave background (CMB) is not responsible for the quick rise the quick rise in main-sequence galaxies. The CMB temperature at z=4 is 13.5\,K. The dust temperature that our high-redshift galaxies would have for a virtually z=0 CMB temperature, $T_{\rm dust}^{z=0}$, is estimated following \citet{Da_Cunha2013}
\begin{equation}
T_{\rm dust}^{z=0} = \left ( (T_{\rm dust}^{\rm meas})^{4+\beta} - (T_{\rm CMB}^{z=0})^{4+\beta} \left [ (1+z)^{4+\beta} -1 \right ] \right )^{\frac{1}{4+\beta}},
\end{equation}
where $T_{\rm CMB}^{z=0}$ is the temperature of the CMB at z=0 and $T_{\rm dust}^{\rm meas}$ is the measured dust temperature at high redshift. This temperature is estimated fitting a gray body with an emissivity of $\beta$=1.8 to our photometric measurements at $\lambda_{\rm rest}>$50\,$\mu$m. The CMB has a relative impact which is lower than 2$\times 10^{-4}$ at all redshifts and thus this effect is negligible. These values are small compared to \citet{Da_Cunha2013}, who assumed a dust temperature of 18\,K. The warmer dust temperatures we measured suggests that the CMB should be less problematic than anticipated.\\

Concerning the evolution of $\langle U \rangle$ in strong starbursts, we found no evidence of evolution ($\propto (1+z)^{-0.1\pm1.0}$) and our results can be fitted by a constant $\langle U \rangle$ of 31$\pm$3. Our value of $\langle U \rangle$ at 0.5$<$z$<$3 is similar to the measurements on a sample of local ULIRGs \citep{Da_Cunha2008b}. This suggests that high-redshift strong starbursts are a more extended version of the nuclei of local ULIRGs, as also suggested by the semi-analytical model of \citet{Lagos2012}. At z$\sim$2.5, the main-sequence galaxies and the strong starbursts have similar $\langle U \rangle$ values. However, we do not interpret the origins of these high values of $\langle U \rangle$ in the same way (see Sect.\,\ref{sect:mdms}, \ref{sect:fgas}, and \ref{discussion}). At z$>2.5$, we cannot constrain with our analysis if $\langle U \rangle$ in strong starbursts rises as in main-sequence galaxies or stays constant.\\

\subsection{Evolution of the ratio between dust and stellar mass}

\label{sect:mdms}

\begin{figure}
\centering
\includegraphics{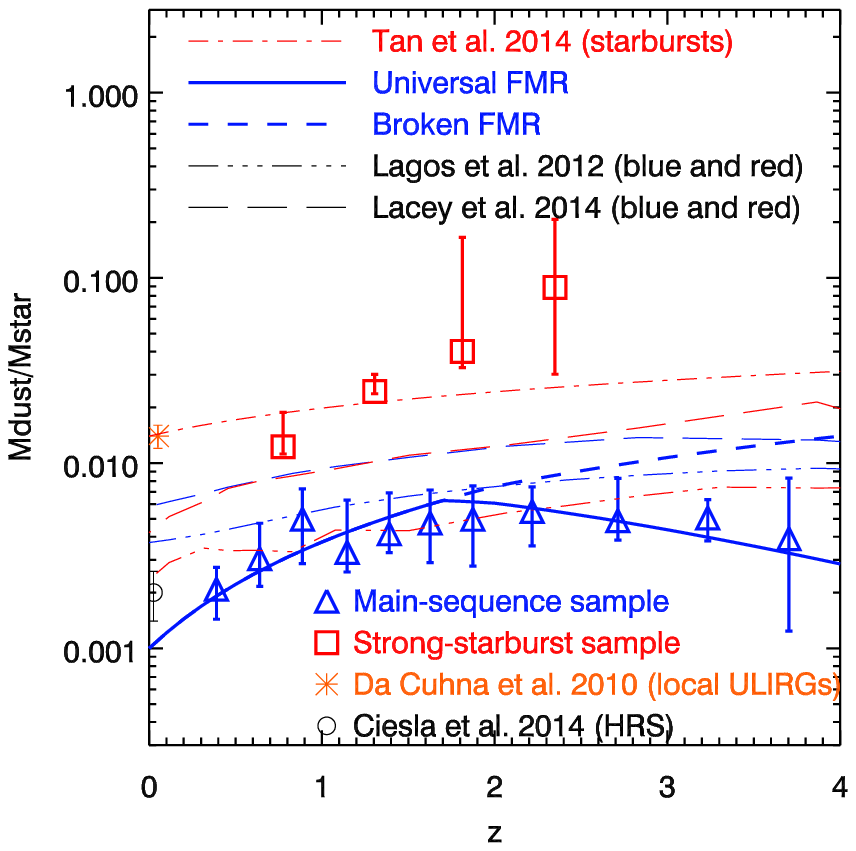}
\caption{\label{fig:MdMs} Mean ratio between dust and stellar mass as a function of redshift in main-sequence galaxies (blue triangles) and strong starbursts (red squares). The orange asterisk is the mean value found for the local ULIRG sample of \citet{Da_Cunha2008b} (see \citealt{Magdis2012b}). The black circle is the average value in HRS galaxies \citep{Ciesla2014}. The solid and dashed lines represent the evolutionary trends expected for a broken and universal FMR, respectively (see Sect.\,\ref{sect:U}). The red dot-dashed line is the best-fit of the evolution found for a sample of individually-detected starbursts of \citet{Tan2014}. The predictions of the models of \citet{Lagos2012} and \citet{Lacey2014} after applying the same mass cut and sSFR selection are overplotted with a three-dot-dash line and a long-dash line, respectively, with the same color code as the symbols.}
\end{figure}

We also studied the evolution of the mean ratio between the dust and the stellar mass in the main-sequence galaxies and the strong starbursts. The results are presented Fig.\,\ref{fig:MdMs}. In main-sequence galaxies, this dust-to-stellar-mass ratio rises up to z$\sim$1 and flattens above this redshift. Strong starbursts typically have 5 times higher ratio. Our measurements are compatible within 2$\sigma$ with the slowly rising trend of $(1+z)^{0.05}$ found by \citet{Tan2014} for a compilation of individual starbursts. However, our data favors a steeper slope.\\

We modeled the evolution of this ratio in main-sequence galaxies using the same simple considerations as in Sect.\,\ref{sect:U}. The evolution of the mean dust-to-stellar-mass ratio can be written as
\begin{equation}
\frac{M_{\rm dust}}{M_\star} \propto \frac{Z(M_\star, \textrm{SFR}) \times M_{\rm mol}}{M_\star} \propto \frac{Z(M_\star, \textrm{SFR}) \times \textrm{SFR}^\beta}{M_\star}.
\end{equation}
One can see that $M_{\rm dust}/M_\star$ is the result of a competition between the rising SFR with increasing redshift and the decreasing gas metallicity. The results are compatible with the broken FMR at 1$\sigma$. The relation obtained with the universal FMR rises too rapidly at high redshift.\\

We also compared our results with predictions of two semi-analytical models. The \citet{Lagos2012} and \citet{Lacey2014} models are based on GALFORM. The main difference between these two models is that \citep{Lagos2012} adopt a universal IMF (a Galactic-like IMF; \citealt{Kennicutt1983}), while \citet{Lacey2014} adopt a non-universal IMF. In the latter star formation taking place in galaxy disks has a Galactic-like IMF, while starbursts have a more top-heavy IMF. This is done to reproduce the number counts of submillimeter galaxies found by surveys.\\

 We select galaxies in the models in the same way we do in the observations based on their stellar mass and distance from the main sequence. An important consideration is that to derive stellar masses in the observations we fix the IMF to a Chabrier IMF, which is different to the IMFs adopted in both models. In order to correct for this we multiply stellar masses in the \citet{Lagos2012} model by 1.1 to go from a Kennicutt IMF to a Chabrier IMF. However, this is non-trivial for the \citet{Lacey2014} model, since it adopts two different IMFs. In order to account for this we correct the fraction of the stellar mass that was formed in the disk by the same factor of 1.1, and divide the fraction of stellar mass that was formed during starbursts by 2. The latter factor is taken as an approximation to go from their adopted top-heavy IMF to a Chabrier IMF, but this conversion is not necessarily unique, and it depends on the dust extinction and stellar age (see \citealt{Mitchell2013} for details). In this paper we make a unique correction, but warn the reader that a more accurate approach would be to perform SED fitting to the predicted SEDs of galaxies and calculating the stellar mass in the same way we would do for observations.\\

Compared to the observations of main-sequence galaxies, the \citet{Lagos2012} model reproduces observations well in the redshift range 1$<$z$<$3, while at $z<1$ and $z>3$ it overpredicts the dust-to-stellar mass ratio. There are different ways to explain the high dust-to-stellar mass ratios: high gas metallicities, high gas masses or stellar masses being too low for the dust masses. In the case of the  \citet{Lagos2012} model the high dust-to-stellar mass ratios are most likely coming from massive galaxies being too gas rich since their metallicities are close to solar, which is what we observe in local galaxies of the same stellar mass range. The \citet{Lacey2014} model predicts dust-to-stellar mass ratios that are twice too high compared to the observations in the whole redshift range. In this case this is because the gas metallicities of MS galaxies in the Lacey model are predicted to be supersolar on average (close to twice the solar metallicity, 12+log(O/H)$\sim$9.0), resulting in dust masses that are higher than observed.\\

In the case of starbursts, the high values inferred for the dust-to-stellar mass ratio in the observations are difficult to interpret. The \citet{Lagos2012} model underpredicts this quantity by a factor of $\sim$5 and the \citet{Lacey2014} model by a factor of $\sim$2. At first the ratio of 1.5-2\% inferred in the observations seems unphysical. However, since the gas fraction (defined here as $\rm M_{\rm mol}/(M_{\rm mol}+M_\star)$) in these high-redshift starbursts is around 50\% (see Sect.\,\ref{sect:fgas}, but also, e.g., \citealt{Riechers2013} and \citealt{Fu2013}), the high values observed for the dust-to-stellar mass ratio can be reached if the gas-to-dust ratio is 50-67. Values similar to the latter are observed in metal-rich galaxies (12+log(O/H)$\sim$9, e.g., \citealt{Remy2014}). This high metal enrichment in strong starbursts compared to main-sequence galaxies could be explained by several mechanisms:
\begin{itemize}
\item the transformation of gas into stars is quicker and the metals are not diluted by the accretion of pristine gas;
\item a fraction of the external layers of low-metallicity gas far from the regions of star formation could be ejected by the strong outflows caused by these extreme starbursts;
\item a top-heavy IMF could produce quickly lots of metals through massive stars without increasing too rapidly the total stellar mass because of mass losses.
\end{itemize}
This high ratio in strong starbursts is discussed in details in \citet{Tan2014}.\\

When it comes to the comparison with the models, one can understand the lower dust-to-stellar mass ratios predicted by the model as resulting from the predicted gas metallicities. \citet{Lagos2012} predict that the average gas metallicity in strong starbursts is close to 0.4 solar metallicities (12+log(O/H)$\sim$8.3), which is about 4 times lower than we can infer from a gas-to-dust mass ratio of $\approx 50$ (see previous paragraph). While the \citet{Lacey2014} model predicts gas metallicities for starbursts that are on average close to solar metallicity (12+log(O/H)$\sim$8.7), 2 times too low for the inferred metallicity of the strong starbursts we observe. We note that both models predict main sequence galaxies having higher metallicities than bright starbursts of the same stellar masses. This seems to contradict the observations and may be at the heart of why the models struggle to get the dust-to-stellar mass ratios of both the main sequence and starburst populations at the same time.

\begin{figure}
\centering
\includegraphics{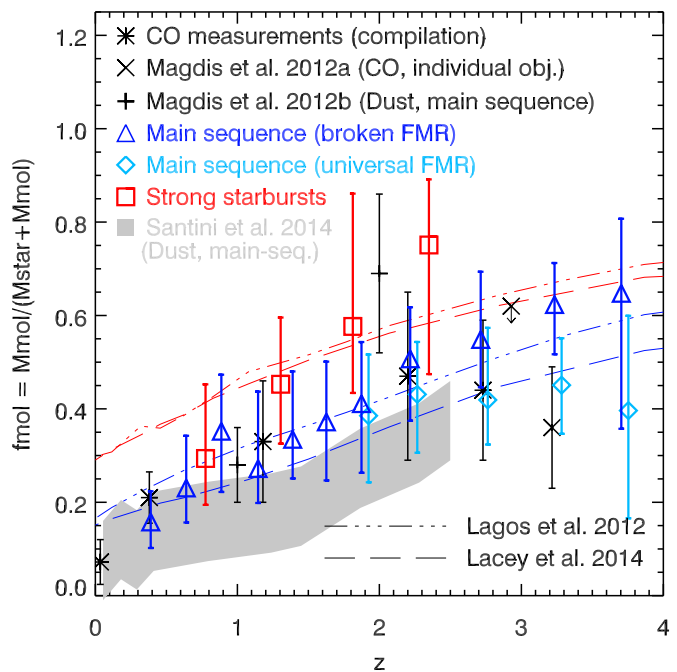}
\caption{\label{fig:gasfrac} Evolution of the mean molecular gas fraction in massive galaxies ($>3\times10^{10}$\,M$_\odot$). The starbursts are represented by red squares and the main-sequence galaxies by blue triangles or light blue diamonds depending on wether the gas fraction is estimated using a broken or an universal FMR, respectively. These results are compared with previous estimate using dust masses of \citet[][black plus]{Magdis2012b} and  \citet[][gray area]{Santini2014}, using CO for two z$>$3 galaxies \citep[][black crosses]{Magdis2012a}, and the compilation of CO measurements of \citet[][black asterisks]{Saintonge2013}. The predictions of the models of \citet{Lagos2012} and \citet{Lacey2014} for the same mass cut are overplotted with a three-dot-dash line and a long-dash line, respectively.}
\end{figure}

\subsection{Evolution of the fraction of molecular gas}

\label{sect:fgas}

Finally, we deduced the mean mass of molecular gas from the dust mass using the same method following \citet{Magdis2011} and \citet{Magdis2012b}. They assumed that the gas-to-dust ratio depends only on gas metallicity and used the local relation of \citet{Leroy2011}\footnote{converted to PP04 convention}:
\begin{equation}
\textrm{log} \left ( \frac{M_{\rm dust}}{M_{\rm mol}} \right ) = (10.54 \pm 1.0) - (0.99\pm0.12) \times {12 + \textrm{log(O/H)}}.
\label{eq:gdr}
\end{equation}
Given the relatively high stellar mass of our samples, and the rising gas masses and ISM pressures to high redshifts \citep{Obreschkow2009}, we expect the contribution of atomic hydrogen to the total gas mass to be negligible and we neglect it in the rest of the paper, considering total gas mass or molecular gas mass to be equivalent. For main-sequence galaxies, the gas metallicity is estimated using the FMR as explained in Sect.\,\ref{sect:U}. We converted the values provided by the FMR from the KD02 to the PP04 metallicity scale using the prescriptions of \citet{Kewley2008} before using it in Eq.\,\ref{eq:gdr}.\\

The gas metallicity in strong starbursts cannot be estimated using the FMR. Indeed, this relation predicts that, at fixed stellar mass, objects forming more stars are less metallic. This effect is expected in gas regulated systems, because a higher accretion of pristine gas involves a stronger SFR, but also a dilution of metals \citep[e.g.,][]{Lilly2013}. This phenomenon is not expected to happen in starbursts, since their high SFRs are not caused by an excess of accretion, but more likely by a major merger. These high-redshift starbursts are probably progenitors of current, massive, elliptical galaxies \citep[e.g.,][]{Toft2014}. We thus assumed that their gas metallicity is similar and used a value of 12+log(O/H) = 9.1$\pm$0.2 (see a detailed discussion in \citealt{Magdis2011} and \citealt{Magdis2012b}).\\

We then derived the molecular gas fraction in main-sequence galaxies, defined in this paper as $\rm M_{mol} / (M_\star + M_{mol})$. The results are presented in Fig.\,\ref{fig:gasfrac}. We found a quick rise up to z$\sim$2. At higher redshifts, the recovered trend depends on the assumptions on the gas metallicity. The rise of the gas fraction in main-sequence galaxies continues at higher redshift if we assume the broken FMR favored by the recent studies, but flattens with a universal FMR. If the broken FMR scenario is confirmed, there could thus be no flattening or reversal of the molecular gas fraction at z$>$2 contrary to what is claimed in \citet{Magdis2012a}, \citet{Saintonge2013}, and \citet{Tan2013}. Our estimations agree with the previous estimates of \citet{Magdis2012b} at z=1, but are 1$\sigma$ lower at z=2, because the bias introduced by clustering was corrected in our study. Our results also agree at 1\,$\sigma$ with the analysis of \citet{Santini2014} at the same stellar mass up to z=2.5 after converting the stellar mass from a Salpeter to a Chabrier IMF convention. However, our estimates are systematically higher than theirs and agree better with the CO data. Our measurements also agree with the compilation of CO measurements of \citet{Saintonge2013} and the two galaxies studied at z$\sim$3 by \citet{Magdis2012a}. These measurements are dependent on the assumed $\alpha_{\rm CO}$ conversion factor, and on the completeness corrections. The good agreement with this independent method is thus an interesting clue to the reliability of these two techniques.\\

Strong starbursts have molecular gas fractions 1$\sigma$ higher than main-sequence galaxies, but follows the same trend. \citet{Sargent2014} predicted that starbursts on average should have a deficit of gas compared to the main sequence (but that gas fraction are expected to rise continuously as the sSFR-excess with respect to the MS increases). Here we selected only the most extreme starbursts with an excess of sSFR of a factor of 10 instead of the average value of $\sim$4. These extreme starbursts may only be possible by the mergers of two gas-rich galaxies galaxies already above the main-sequence before the merger. This could explain this small positive offset compared to the main-sequence sample.\\

We also compared our results with the models of \citet{Lagos2012} and \citet{Lacey2014} presented in Sect.\,\ref{sect:mdms}. Both models agree well with our measurements of the gas fraction for starburst galaxies at all redshifts and main-sequence galaxies at 1.5$<$z$<$3. Both the \citet{Lagos2012} and \citet{Lacey2014} models overpredict the molecular gas fraction at z$<$0.5 at a 1-2$\sigma$ level. At reshifts $z>3$, the \citet{Lacey2014} model agrees better with the universal FMR scenario at z$>$3, while the \citet{Lagos2012} model is more compatible with the broken FMR. The fact that both models predict molecular gas fractions that in overall agree with the observations supports our interpretation in Sect.~\ref{sect:mdms}, which points to the model of metal enrichment as the source of discrepancy in the dust-to-stellar mass ratios.\\ 

\subsection{Evolution of the depletion time}

\begin{figure}
\centering
\includegraphics{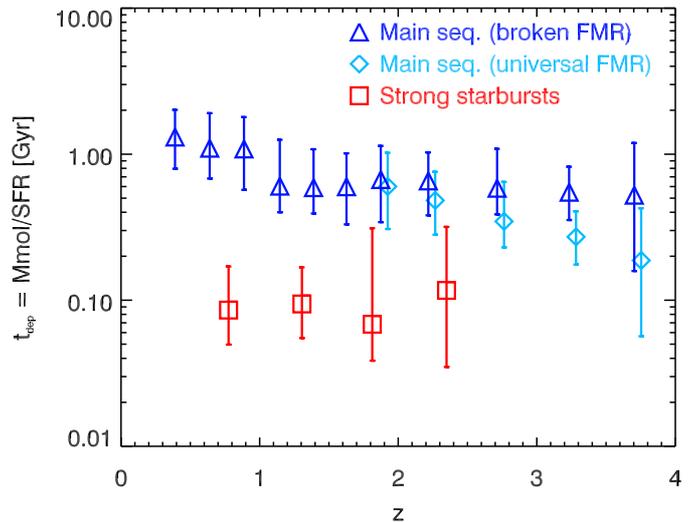}
\caption{\label{fig:tdep} Evolution of the mean molecular gas depletion time. The symbols are the same as in Fig.\,\ref{fig:gasfrac}.}
\end{figure}

We estimated the mean depletion time of the molecular gas, defined in our analysis as the ratio between the mass of molecular gas and the SFR. Figure\,\ref{fig:tdep} shows our results. The depletion time in strong starbursts does not evolve with redshift and is compatible with 100\,Myr, the typical timescale of the strong boost of star formation induced by major mergers \citep[e.g.,][]{Di_Matteo2008}. This timescale is longer in main-sequence galaxies and slightly (1\,$\sigma$) evolves with redshift at z$<$1. It decreases from 1.3$_{-0.5}^{+0.7}$\,Gyr at z$\sim$0.375 to $\sim$500\,Myr around z$\sim$1.5 and is stable at higher redshift in the case of a broken FMR (but continues to decrease with redshift for a universal FMR). This timescale is similar to the maximum duration high-redshift massive galaxies can stay on the main-sequence before reaching the quenching mass around 10$^{11}$\,M$_\odot$ \citep{Heinis2014}. The mass of molecular gas and stars contained in these high-redshift objects is already sufficient to reach this quenching mass without any additional accretion of gas.\\

\section{Discussion}

\label{discussion}

\subsection{What is the main driver of the strong evolution of the specific star formation rate?}

\begin{figure}
\centering
\includegraphics{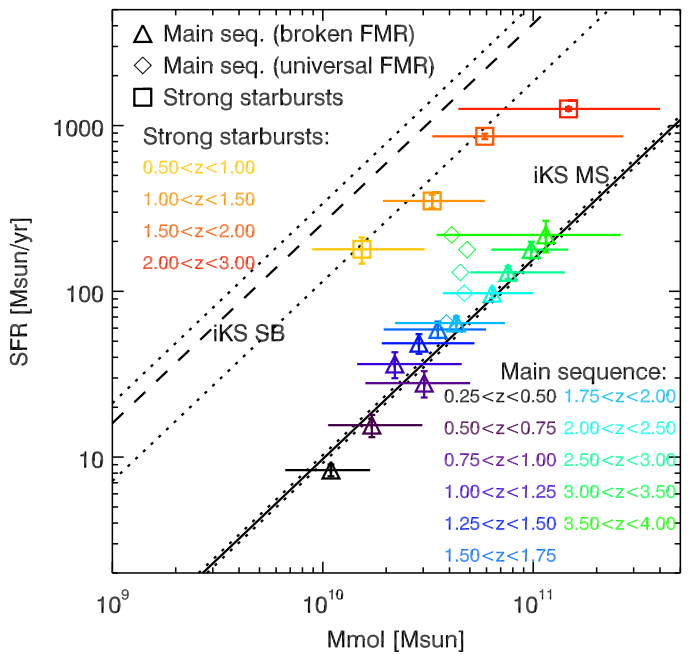}
\caption{\label{fig:iKS} Relation between the mean SFR rate and the mean molecular gas mass in our galaxy samples, i.e., integrated Kennicutt-Schmidt law. The solid line and the dashed line are the center of the relation fitted by \citet{Sargent2014} on a compilation of data for main-sequence galaxies and starbursts, respectively. The dotted lines represent the 1\,$\sigma$ uncertainties on these relations.} The triangles and diamonds represent the average position of massive, main-sequence galaxies in this diagram assuming a broken FMR and an universal FMR, respectively. The squares indicates the average position of strong starbursts.
\end{figure}

We checked the average position of our selection of massive galaxies in the integrated Kennicutt-Schmidt diagram (SFR versus mass of molecular gas) to gain insight on their mode of star formation. In this diagram, normal star-forming galaxies and starbursts follow two distinct sequences. For comparison, we used the fit of a recent data compilation performed by \citet{Sargent2014}. The results are presented in Fig.\,\ref{fig:iKS}.\\

The average position of our sample of strong starbursts is in the 1$\sigma$ confidence region of \citet{Sargent2014} for starbursts. They are systematically below the central relation, but the uncertainty is dominated by the systematic uncertainties on their gas metallicity. In addition, \citet{Sargent2014} suggested that the SFEs of starbursts follow a continuum of values depending on their boost of sSFR. Our objects are thus not expected to be exactly on the central relation. The interpretation of the results for main-sequence galaxies is dependent on the hypothesis on the gas metallicity. In the scenario of a broken FMR favored by recent observations, the average position of main-sequence galaxies at all redshifts falls on the relation of normal star-forming galaxies. This suggests that the star formation is dominated by galaxies forming their stars through a normal mode at all redshifts below z=4. In the case of a non-evolving FMR, the massive high-redshift galaxies do not stay on the normal star-forming sequence and have higher SFEs.\\

If the scenario of a broken FMR favored by the most recent observations is consolidated, the strong star-formation observed in massive high-redshift galaxies would thus be caused by huge gas reservoirs probably fed by an intense cosmological accretion. This strong accretion of primordial gas dilute the metals produced by the massive stars \citep[e.g.,][]{Bouche2010,Lilly2013}. Consequently, the gas-to-dust ratio is much lower at high redshift than at low redshift. Since the star-formation efficiency is only slowly evolving (SFR$\propto$M$_{\rm mol}^{1.2}$), the number of UV photons absorbed per mass of dust is thus higher and the dust temperature is warmer as observed in our analysis (see Sect.\,\ref{sect:U}). This scenario provides thus a consistent interpretation of evolution of both the sSFR and the dust temperature of massive galaxies with redshift.\\

\subsection{Limitations of our approach}

Our analysis provided suggestive results. However, it relies on several hypotheses, which cannot be extensively tested yet. In this section, we discuss the potential limitation of our analysis.\\

The evolution of the metallicity relations at z$>$2.5 was measured only by a few pioneering works, which found that the normalization of the FMR evolves at z$>$2.5. We used a simple renormalization depending on redshift to take this evolution into account. The redshift sampling of these studies is relatively coarse and we used a simple linear evolution with redshift. Future studies based on larger samples will allow a finer sampling of the evolution of the gas metallicity in massive galaxies at high redshift. However, the current results are very encouraging. The current assumption of a broken FMR allows us to recover naturally both the evolution of the $\langle U \rangle$ parameter and the integrated Schmidt-Kennicutt relation at high redshift.\\

The gas metallicity of strong starbursts was more problematic to set. We can reasonably guess it assuming they are progenitors of the most massive galaxies. However, direct measurements of their gas metallicity are difficult to perform using optical/near-IR spectroscopy because of their strong dust attenuation. The millimeter spectroscopy of fine-structure lines with ALMA will be certainly an interesting way to determine the distribution of gas metallicity of strong starbursts in the future \citep[e.g.,][]{Nagao2011}.\\

The validity of the calibration of the gas-to-dust ratio versus gas metallicity relation in most extreme environment is also uncertain and difficult to test with the current data sets. \citet{Saintonge2013} found an offset of a factor 1.7 for a population of lensed galaxies and discussed the possible origins of the tension between the gas content estimated from CO and from dust. However, we found no offset with the integrated Kennicutt-Schmidt relation in our analysis and a good agreement with the compilation of CO measurements of gas fractions. The lensed galaxies of \citet{Saintonge2013} could be a peculiar population because they are UV-selected and then biased toward dust-poor systems. They could also be affected differential magnification effects or \textit{Herschel}-selection biases. The hypotheses performed to estimate the gas metallicity are also different between their and our analysis (standard mass-metallicity relation versus broken FMR).\\

Finally, the stacking analysis only provides an average measurement of a full population. Thus it is difficult to estimate the heterogeneity of the stacked populations. Bootstrap techniques can be applied to estimate the scatter on the flux density at a given wavelength \citep{Bethermin2012b}. However, because of the correlation between $\langle U \rangle$ and M$_d$, this technique cannot be applied to measure the scatter on each of these parameters.\\

\section{Conclusion}

We used a stacking analysis to measure the evolution of the average mid-infrared to millimeter emission of massive star-forming galaxies up to z=4. We then derived the evolution of the mean physical parameters of these objects. Our main findings are the following.

\begin{itemize}
\item The mean intensity of the radiation field $\langle U \rangle$ in main-sequence galaxies, which is strongly correlated with their dust temperature, rises rapidly with redshift: $\langle U \rangle = (3.0\pm1.1) \times (1+z)^{1.8 \pm 0.4}$. This evolution can be interpreted considering the decrease in the gas metallicity of galaxies at constant stellar mass with increasing redshift. We found no evidence for an evolution of $\langle U \rangle$ in strong starbursts up to z=3.\\
\item The mean ratio between the dust mass and the stellar mass in main-sequence galaxies rises between z=0 and z=1 and exhibit a plateau at higher redshift. The strong starbursts have a higher ratio by a factor of 5.\\
\item The average fraction of molecular gas ($\rm M_{mol} / (M_\star + M_{mol})$) rises rapidly with redshift and reaches $\sim$60\% at z=4. A similar evolution is found in strong starbursts, but with slightly higher values. These results agree with the pilot CO surveys performed at high redshift.\\
\item We compare with two state-of-the-art semi-analytic models that adopt either a universal IMF or a top-heavy IMF in starbursts and find that the models predict molecular gas fractions that agree well with the observations but the predicted dust-to-stellar mass ratios are either too high or too low. We interpret this as being due to the way metal enrichment is dealt with in the simulations. We suggest different mechanisms that can help overcome this issue. For instance, outflows affecting more metal depleted gas that is in the outer parts of galaxies.\\ 
\item The average position of the massive main-sequence galaxies in the integrated Kennicutt-Schmidt diagram corresponds to the sequence of normal star-forming galaxies. This suggests that the bulk of the star-formation up to z$\sim$4 is dominated by the normal mode of star-formation and that the extreme SFR observed are caused by huge gas reservoirs probably induced by the very intense cosmological accretion. The strong starbursts follow another sequence with a 5--10 times higher star-formation efficiency.\\

\end{itemize}

\begin{acknowledgements}
We thank the anonymous referee for providing constructive comments. We acknowledge Morgane Cousin, Nick Lee, Nick Scoville, and Christian Maier for their interesting discussions/suggestions, Laure Ciesla for providing an electronic table of the physical properties of the HRS sample, and Amélie Saintonge for providing her compilation of data. We gratefully acknowledge the contributions of the entire COSMOS collaboration consisting of more than 100 scientists. The HST COSMOS program was supported through NASA grant HST-GO-09822. More information on the COSMOS survey is available at \url{http://www.astro.caltech.edu/cosmos}. ased on data obtained from the ESO Science Archive Facility. Based on data products from observations made with ESO Telescopes at the La Silla Paranal Observatory under ESO programme ID 179.A-2005 and on data products produced by TERAPIX and the Cambridge Astronomy Survey Unit on behalf of the UltraVISTA consortium. MB, ED, and MS acknowledge the support of the ERC-StG UPGAL 240039 and ANR-08-JCJC-0008 grants. AK acknowledges support by the Collaborative Research Council 956, sub-project A1, funded by the Deutsche Forschungsgemeinschaft (DFG).


\end{acknowledgements}

\bibliographystyle{aa}

\bibliography{biblio}

\begin{thebibliography}{108}
\expandafter\ifx\csname natexlab\endcsname\relax\def\natexlab#1{#1}\fi

\bibitem[{{Amor{\'{\i}}n} {et~al.}(2014){Amor{\'{\i}}n}, {Grazian},
  {Castellano}, {Pentericci}, {Fontana}, {Sommariva}, {van der Wel}, {Maseda},
  \& {Merlin}}]{Amorin2014}
{Amor{\'{\i}}n}, R., {Grazian}, A., {Castellano}, M., {et~al.} 2014, \apjl,
  788, L4

\bibitem[{{Aravena} {et~al.}(2013){Aravena}, {Murphy}, {Aguirre}, {Ashby},
  {Benson}, {Bothwell}, {Brodwin}, {Carlstrom}, {Chapman}, {Crawford}, {de
  Breuck}, {Fassnacht}, {Gonzalez}, {Greve}, {Gullberg}, {Hezaveh}, {Holder},
  {Holzapfel}, {Keisler}, {Malkan}, {Marrone}, {McIntyre}, {Reichardt},
  {Sharon}, {Spilker}, {Stalder}, {Stark}, {Vieira}, \&
  {Wei{\ss}}}]{Aravena2013}
{Aravena}, M., {Murphy}, E.~J., {Aguirre}, J.~E., {et~al.} 2013, \mnras, 433,
  498

\bibitem[{{Aretxaga} {et~al.}(2011){Aretxaga}, {Wilson}, {Aguilar}, {Alberts},
  {Scott}, {Scoville}, {Yun}, {Austermann}, {Downes}, {Ezawa}, {Hatsukade},
  {Hughes}, {Kawabe}, {Kohno}, {Oshima}, {Perera}, {Tamura}, \&
  {Zeballos}}]{Aretxaga2011}
{Aretxaga}, I., {Wilson}, G.~W., {Aguilar}, E., {et~al.} 2011, \mnras, 415,
  3831

\bibitem[{{Arnouts} {et~al.}(1999){Arnouts}, {Cristiani}, {Moscardini},
  {Matarrese}, {Lucchin}, {Fontana}, \& {Giallongo}}]{Arnouts1999}
{Arnouts}, S., {Cristiani}, S., {Moscardini}, L., {et~al.} 1999, \mnras, 310,
  540

\bibitem[{{Bavouzet}(2008)}]{Bavouzet2008}
{Bavouzet}, N. 2008, PhD thesis, Universit\'e Paris-Sud 11

\bibitem[{{Behroozi} {et~al.}(2010){Behroozi}, {Conroy}, \&
  {Wechsler}}]{Behroozi2010}
{Behroozi}, P.~S., {Conroy}, C., \& {Wechsler}, R.~H. 2010, \apj, 717, 379

\bibitem[{{B{\'e}thermin} {et~al.}(2012{\natexlab{a}}){B{\'e}thermin}, {Daddi},
  {Magdis}, {Sargent}, {Hezaveh}, {Elbaz}, {Le Borgne}, {Mullaney}, {Pannella},
  {Buat}, {Charmandaris}, {Lagache}, \& {Scott}}]{Bethermin2012c}
{B{\'e}thermin}, M., {Daddi}, E., {Magdis}, G., {et~al.} 2012{\natexlab{a}},
  \apjl, 757, L23

\bibitem[{{B{\'e}thermin} {et~al.}(2010{\natexlab{a}}){B{\'e}thermin}, {Dole},
  {Beelen}, \& {Aussel}}]{Bethermin2010a}
{B{\'e}thermin}, M., {Dole}, H., {Beelen}, A., \& {Aussel}, H.
  2010{\natexlab{a}}, \aap, 512, A78+

\bibitem[{{B{\'e}thermin} {et~al.}(2010{\natexlab{b}}){B{\'e}thermin}, {Dole},
  {Cousin}, \& {Bavouzet}}]{Bethermin2010b}
{B{\'e}thermin}, M., {Dole}, H., {Cousin}, M., \& {Bavouzet}, N.
  2010{\natexlab{b}}, \aap, 516, A43+

\bibitem[{{B{\'e}thermin} {et~al.}(2011){B{\'e}thermin}, {Dole}, {Lagache}, {Le
  Borgne}, \& {Penin}}]{Bethermin2011}
{B{\'e}thermin}, M., {Dole}, H., {Lagache}, G., {Le Borgne}, D., \& {Penin}, A.
  2011, \aap, 529, A4+

\bibitem[{{B{\'e}thermin} {et~al.}(2014){B{\'e}thermin}, {Kilbinger}, {Daddi},
  {Gabor}, {Finoguenov}, {McCracken}, {Wolk}, {Aussel}, {Strazzulo}, {Le
  Floc'h}, {Gobat}, {Rodighiero}, {Dickinson}, {Wang}, {Lutz}, \&
  {Heinis}}]{Bethermin2014}
{B{\'e}thermin}, M., {Kilbinger}, M., {Daddi}, E., {et~al.} 2014, \aap, 567,
  A103

\bibitem[{{B{\'e}thermin} {et~al.}(2012{\natexlab{b}}){B{\'e}thermin}, {Le
  Floc'h}, {Ilbert}, {Conley}, {Lagache}, {Amblard}, {Arumugam}, {Aussel},
  {Berta}, {Bock}, {Boselli}, {Buat}, {Casey}, {Castro-Rodr{\'{\i}}guez},
  {Cava}, {Clements}, {Cooray}, {Dowell}, {Eales}, {Farrah}, {Franceschini},
  {Glenn}, {Griffin}, {Hatziminaoglou}, {Heinis}, {Ibar}, {Ivison},
  {Kartaltepe}, {Levenson}, {Magdis}, {Marchetti}, {Marsden}, {Nguyen},
  {O'Halloran}, {Oliver}, {Omont}, {Page}, {Panuzzo}, {Papageorgiou},
  {Pearson}, {P{\'e}rez-Fournon}, {Pohlen}, {Rigopoulou}, {Roseboom},
  {Rowan-Robinson}, {Salvato}, {Schulz}, {Scott}, {Seymour}, {Shupe}, {Smith},
  {Symeonidis}, {Trichas}, {Tugwell}, {Vaccari}, {Valtchanov}, {Vieira},
  {Viero}, {Wang}, {Xu}, \& {Zemcov}}]{Bethermin2012b}
{B{\'e}thermin}, M., {Le Floc'h}, E., {Ilbert}, O., {et~al.}
  2012{\natexlab{b}}, \aap, 542, A58

\bibitem[{{B{\'e}thermin} {et~al.}(2013){B{\'e}thermin}, {Wang}, {Dor{\'e}},
  {Lagache}, {Sargent}, {Daddi}, {Cousin}, \& {Aussel}}]{Bethermin2013}
{B{\'e}thermin}, M., {Wang}, L., {Dor{\'e}}, O., {et~al.} 2013, \aap, 557, A66

\bibitem[{{Bothwell} {et~al.}(2010){Bothwell}, {Chapman}, {Tacconi}, {Smail},
  {Ivison}, {Casey}, {Bertoldi}, {Beswick}, {Biggs}, {Blain}, {Cox}, {Genzel},
  {Greve}, {Kennicutt}, {Muxlow}, {Neri}, \& {Omont}}]{Bothwell2010}
{Bothwell}, M.~S., {Chapman}, S.~C., {Tacconi}, L., {et~al.} 2010, \mnras, 405,
  219

\bibitem[{{Bouch{\'e}} {et~al.}(2010){Bouch{\'e}}, {Dekel}, {Genzel}, {Genel},
  {Cresci}, {F{\"o}rster Schreiber}, {Shapiro}, {Davies}, \&
  {Tacconi}}]{Bouche2010}
{Bouch{\'e}}, N., {Dekel}, A., {Genzel}, R., {et~al.} 2010, \apj, 718, 1001

\bibitem[{{Bournaud} {et~al.}(2007){Bournaud}, {Elmegreen}, \&
  {Elmegreen}}]{Bournaud2007}
{Bournaud}, F., {Elmegreen}, B.~G., \& {Elmegreen}, D.~M. 2007, \apj, 670, 237

\bibitem[{{Bourne} {et~al.}(2012){Bourne}, {Maddox}, {Dunne}, {Auld}, {Baes},
  {Baldry}, {Bonfield}, {Cooray}, {Croom}, {Dariush}, {de Zotti}, {Driver},
  {Dye}, {Eales}, {Gomez}, {Gonz{\'a}lez-Nuevo}, {Hopkins}, {Ibar}, {Jarvis},
  {Lapi}, {Madore}, {Micha{\l}owski}, {Pohlen}, {Popescu}, {Rigby}, {Seibert},
  {Smith}, {Tuffs}, {Werf}, {Brough}, {Buttiglione}, {Cava}, {Clements},
  {Conselice}, {Fritz}, {Hopwood}, {Ivison}, {Jones}, {Kelvin}, {Liske},
  {Loveday}, {Norberg}, {Robotham}, {Rodighiero}, \& {Temi}}]{Bourne2012}
{Bourne}, N., {Maddox}, S.~J., {Dunne}, L., {et~al.} 2012, \mnras, 421, 3027

\bibitem[{{Burgarella} {et~al.}(2013){Burgarella}, {Buat}, {Gruppioni},
  {Cucciati}, {Heinis}, {Berta}, {B{\'e}thermin}, {Bock}, {Cooray}, {Dunlop},
  {Farrah}, {Franceschini}, {Le Floc'h}, {Lutz}, {Magnelli}, {Nordon},
  {Oliver}, {Page}, {Popesso}, {Pozzi}, {Riguccini}, {Vaccari}, \&
  {Viero}}]{Burgarella2013}
{Burgarella}, D., {Buat}, V., {Gruppioni}, C., {et~al.} 2013, \aap, 554, A70

\bibitem[{{Chabrier}(2003)}]{Chabrier2003}
{Chabrier}, G. 2003, \pasp, 115, 763

\bibitem[{{Ciesla} {et~al.}(2014){Ciesla}, {Boquien}, {Boselli}, {Buat},
  {Cortese}, {Bendo}, {Heinis}, {Galametz}, {Eales}, {Smith}, {Baes},
  {Bianchi}, {de Looze}, {di Serego Alighieri}, {Galliano}, {Hughes}, {Madden},
  {Pierini}, {R{\'e}my-Ruyer}, {Spinoglio}, {Vaccari}, {Viaene}, \&
  {Vlahakis}}]{Ciesla2014}
{Ciesla}, L., {Boquien}, M., {Boselli}, A., {et~al.} 2014, \aap, 565, A128

\bibitem[{{da Cunha} {et~al.}(2008){da Cunha}, {Charlot}, \&
  {Elbaz}}]{Da_Cunha2008b}
{da Cunha}, E., {Charlot}, S., \& {Elbaz}, D. 2008, \mnras, 388, 1595

\bibitem[{{da Cunha} {et~al.}(2013){da Cunha}, {Groves}, {Walter}, {Decarli},
  {Weiss}, {Bertoldi}, {Carilli}, {Daddi}, {Elbaz}, {Ivison}, {Maiolino},
  {Riechers}, {Rix}, {Sargent}, \& {Smail}}]{Da_Cunha2013}
{da Cunha}, E., {Groves}, B., {Walter}, F., {et~al.} 2013, \apj, 766, 13

\bibitem[{{Daddi} {et~al.}(2010{\natexlab{a}}){Daddi}, {Bournaud}, {Walter},
  {Dannerbauer}, {Carilli}, {Dickinson}, {Elbaz}, {Morrison}, {Riechers},
  {Onodera}, {Salmi}, {Krips}, \& {Stern}}]{Daddi2010a}
{Daddi}, E., {Bournaud}, F., {Walter}, F., {et~al.} 2010{\natexlab{a}}, \apj,
  713, 686

\bibitem[{{Daddi} {et~al.}(2008){Daddi}, {Dannerbauer}, {Elbaz}, {Dickinson},
  {Morrison}, {Stern}, \& {Ravindranath}}]{Daddi2008}
{Daddi}, E., {Dannerbauer}, H., {Elbaz}, D., {et~al.} 2008, \apjl, 673, L21

\bibitem[{{Daddi} {et~al.}(2009{\natexlab{a}}){Daddi}, {Dannerbauer}, {Krips},
  {Walter}, {Dickinson}, {Elbaz}, \& {Morrison}}]{Daddi2009a}
{Daddi}, E., {Dannerbauer}, H., {Krips}, M., {et~al.} 2009{\natexlab{a}},
  \apjl, 695, L176

\bibitem[{{Daddi} {et~al.}(2009{\natexlab{b}}){Daddi}, {Dannerbauer}, {Stern},
  {Dickinson}, {Morrison}, {Elbaz}, {Giavalisco}, {Mancini}, {Pope}, \&
  {Spinrad}}]{Daddi2009b}
{Daddi}, E., {Dannerbauer}, H., {Stern}, D., {et~al.} 2009{\natexlab{b}}, \apj,
  694, 1517

\bibitem[{{Daddi} {et~al.}(2007){Daddi}, {Dickinson}, {Morrison}, {Chary},
  {Cimatti}, {Elbaz}, {Frayer}, {Renzini}, {Pope}, {Alexander}, {Bauer},
  {Giavalisco}, {Huynh}, {Kurk}, \& {Mignoli}}]{Daddi2007}
{Daddi}, E., {Dickinson}, M., {Morrison}, G., {et~al.} 2007, \apj, 670, 156

\bibitem[{{Daddi} {et~al.}(2010{\natexlab{b}}){Daddi}, {Elbaz}, {Walter},
  {Bournaud}, {Salmi}, {Carilli}, {Dannerbauer}, {Dickinson}, {Monaco}, \&
  {Riechers}}]{Daddi2010b}
{Daddi}, E., {Elbaz}, D., {Walter}, F., {et~al.} 2010{\natexlab{b}}, \apjl,
  714, L118

\bibitem[{{Di Matteo} {et~al.}(2008){Di Matteo}, {Bournaud}, {Martig},
  {Combes}, {Melchior}, \& {Semelin}}]{Di_Matteo2008}
{Di Matteo}, P., {Bournaud}, F., {Martig}, M., {et~al.} 2008, \aap, 492, 31

\bibitem[{{Dole} {et~al.}(2003){Dole}, {Lagache}, \& {Puget}}]{Dole2003}
{Dole}, H., {Lagache}, G., \& {Puget}, J. 2003, \apj, 585, 617

\bibitem[{{Draine} \& {Li}(2007)}]{Draine2007}
{Draine}, B.~T. \& {Li}, A. 2007, \apj, 657, 810

\bibitem[{{Elbaz} {et~al.}(2007){Elbaz}, {Daddi}, {Le Borgne}, {Dickinson},
  {Alexander}, {Chary}, {Starck}, {Brandt}, {Kitzbichler}, {MacDonald},
  {Nonino}, {Popesso}, {Stern}, \& {Vanzella}}]{Elbaz2007}
{Elbaz}, D., {Daddi}, E., {Le Borgne}, D., {et~al.} 2007, \aap, 468, 33

\bibitem[{{Elbaz} {et~al.}(2011){Elbaz}, {Dickinson}, {Hwang},
  {D{\'{\i}}az-Santos}, {Magdis}, {Magnelli}, {Le Borgne}, {Galliano},
  {Pannella}, {Chanial}, {Armus}, {Charmandaris}, {Daddi}, {Aussel}, {Popesso},
  {Kartaltepe}, {Altieri}, {Valtchanov}, {Coia}, {Dannerbauer}, {Dasyra},
  {Leiton}, {Mazzarella}, {Alexander}, {Buat}, {Burgarella}, {Chary}, {Gilli},
  {Ivison}, {Juneau}, {Le Floc'h}, {Lutz}, {Morrison}, {Mullaney}, {Murphy},
  {Pope}, {Scott}, {Brodwin}, {Calzetti}, {Cesarsky}, {Charlot}, {Dole},
  {Eisenhardt}, {Ferguson}, {F{\"o}rster Schreiber}, {Frayer}, {Giavalisco},
  {Huynh}, {Koekemoer}, {Papovich}, {Reddy}, {Surace}, {Teplitz}, {Yun}, \&
  {Wilson}}]{Elbaz2011}
{Elbaz}, D., {Dickinson}, M., {Hwang}, H.~S., {et~al.} 2011, \aap, 533, A119

\bibitem[{{Engel} {et~al.}(2010){Engel}, {Tacconi}, {Davies}, {Neri}, {Smail},
  {Chapman}, {Genzel}, {Cox}, {Greve}, {Ivison}, {Blain}, {Bertoldi}, \&
  {Omont}}]{Engel2010}
{Engel}, H., {Tacconi}, L.~J., {Davies}, R.~I., {et~al.} 2010, \apj, 724, 233

\bibitem[{{Frayer} {et~al.}(2008){Frayer}, {Koda}, {Pope}, {Huynh}, {Chary},
  {Scott}, {Dickinson}, {Bock}, {Carpenter}, {Hawkins}, {Hodges}, {Lamb},
  {Plambeck}, {Pound}, {Scott}, {Scoville}, \& {Woody}}]{Frayer2008}
{Frayer}, D.~T., {Koda}, J., {Pope}, A., {et~al.} 2008, \apjl, 680, L21

\bibitem[{{Fu} {et~al.}(2013){Fu}, {Cooray}, {Feruglio}, {Ivison}, {Riechers},
  {Gurwell}, {Bussmann}, {Harris}, {Altieri}, {Aussel}, {Baker}, {Bock},
  {Boylan-Kolchin}, {Bridge}, {Calanog}, {Casey}, {Cava}, {Chapman},
  {Clements}, {Conley}, {Cox}, {Farrah}, {Frayer}, {Hopwood}, {Jia}, {Magdis},
  {Marsden}, {Mart{\'{\i}}nez-Navajas}, {Negrello}, {Neri}, {Oliver}, {Omont},
  {Page}, {P{\'e}rez-Fournon}, {Schulz}, {Scott}, {Smith}, {Vaccari},
  {Valtchanov}, {Vieira}, {Viero}, {Wang}, {Wardlow}, \& {Zemcov}}]{Fu2013}
{Fu}, H., {Cooray}, A., {Feruglio}, C., {et~al.} 2013, \nat, 498, 338

\bibitem[{{Genzel} {et~al.}(2012){Genzel}, {Tacconi}, {Combes}, {Bolatto},
  {Neri}, {Sternberg}, {Cooper}, {Bouch{\'e}}, {Bournaud}, {Burkert},
  {Comerford}, {Cox}, {Davis}, {F{\"o}rster Schreiber}, {Garcia-Burillo},
  {Gracia-Carpio}, {Lutz}, {Naab}, {Newman}, {Saintonge}, {Shapiro}, {Shapley},
  \& {Weiner}}]{Genzel2012}
{Genzel}, R., {Tacconi}, L.~J., {Combes}, F., {et~al.} 2012, \apj, 746, 69

\bibitem[{{Genzel} {et~al.}(2006){Genzel}, {Tacconi}, {Eisenhauer},
  {F{\"o}rster Schreiber}, {Cimatti}, {Daddi}, {Bouch{\'e}}, {Davies},
  {Lehnert}, {Lutz}, {Nesvadba}, {Verma}, {Abuter}, {Shapiro}, {Sternberg},
  {Renzini}, {Kong}, {Arimoto}, \& {Mignoli}}]{Genzel2006}
{Genzel}, R., {Tacconi}, L.~J., {Eisenhauer}, F., {et~al.} 2006, \nat, 442, 786

\bibitem[{{Genzel} {et~al.}(2010){Genzel}, {Tacconi}, {Gracia-Carpio},
  {Sternberg}, {Cooper}, {Shapiro}, {Bolatto}, {Bouch{\'e}}, {Bournaud},
  {Burkert}, {Combes}, {Comerford}, {Cox}, {Davis}, {Schreiber},
  {Garcia-Burillo}, {Lutz}, {Naab}, {Neri}, {Omont}, {Shapley}, \&
  {Weiner}}]{Genzel2010}
{Genzel}, R., {Tacconi}, L.~J., {Gracia-Carpio}, J., {et~al.} 2010, \mnras,
  407, 2091

\bibitem[{{Genzel} {et~al.}(2014){Genzel}, {Tacconi}, {Lutz}, {Saintonge},
  {Berta}, {Magnelli}, {Combes}, {Garc{\'{\i}}a-Burillo}, {Neri}, {Bolatto},
  {Contini}, {Lilly}, {Boissier}, {Boone}, {Bouch{\'e}}, {Bournaud}, {Burkert},
  {Carollo}, {Colina}, {Cooper}, {Cox}, {Feruglio}, {F{\"o}rster Schreiber},
  {Freundlich}, {Gracia-Carpio}, {Juneau}, {Kovac}, {Lippa}, {Naab}, {Salome},
  {Renzini}, {Sternberg}, {Walter}, {Weiner}, {Weiss}, \& {Wuyts}}]{Genzel2014}
{Genzel}, R., {Tacconi}, L.~J., {Lutz}, D., {et~al.} 2014, arXiv:1409.1171

\bibitem[{{Greve} {et~al.}(2005){Greve}, {Bertoldi}, {Smail}, {Neri},
  {Chapman}, {Blain}, {Ivison}, {Genzel}, {Omont}, {Cox}, {Tacconi}, \&
  {Kneib}}]{Greve2005}
{Greve}, T.~R., {Bertoldi}, F., {Smail}, I., {et~al.} 2005, \mnras, 359, 1165

\bibitem[{{Griffin} {et~al.}(2010){Griffin}, {Abergel}, {Abreu}, {Ade},
  {Andr{\'e}}, {Augueres}, {Babbedge}, {Bae}, {Baillie}, {Baluteau}, {Barlow},
  {Bendo}, {Benielli}, {Bock}, {Bonhomme}, {Brisbin}, {Brockley-Blatt},
  {Caldwell}, {Cara}, {Castro-Rodriguez}, {Cerulli}, {Chanial}, {Chen},
  {Clark}, {Clements}, {Clerc}, {Coker}, {Communal}, {Conversi}, {Cox},
  {Crumb}, {Cunningham}, {Daly}, {Davis}, {de Antoni}, {Delderfield}, {Devin},
  {di Giorgio}, {Didschuns}, {Dohlen}, {Donati}, {Dowell}, {Dowell}, {Duband},
  {Dumaye}, {Emery}, {Ferlet}, {Ferrand}, {Fontignie}, {Fox}, {Franceschini},
  {Frerking}, {Fulton}, {Garcia}, {Gastaud}, {Gear}, {Glenn}, {Goizel},
  {Griffin}, {Grundy}, {Guest}, {Guillemet}, {Hargrave}, {Harwit}, {Hastings},
  {Hatziminaoglou}, {Herman}, {Hinde}, {Hristov}, {Huang}, {Imhof}, {Isaak},
  {Israelsson}, {Ivison}, {Jennings}, {Kiernan}, {King}, {Lange}, {Latter},
  {Laurent}, {Laurent}, {Leeks}, {Lellouch}, {Levenson}, {Li}, {Li},
  {Lilienthal}, {Lim}, {Liu}, {Lu}, {Madden}, {Mainetti}, {Marliani}, {McKay},
  {Mercier}, {Molinari}, {Morris}, {Moseley}, {Mulder}, {Mur}, {Naylor},
  {Nguyen}, {O'Halloran}, {Oliver}, {Olofsson}, {Olofsson}, {Orfei}, {Page},
  {Pain}, {Panuzzo}, {Papageorgiou}, {Parks}, {Parr-Burman}, {Pearce},
  {Pearson}, {P{\'e}rez-Fournon}, {Pinsard}, {Pisano}, {Podosek}, {Pohlen},
  {Polehampton}, {Pouliquen}, {Rigopoulou}, {Rizzo}, {Roseboom}, {Roussel},
  {Rowan-Robinson}, {Rownd}, {Saraceno}, {Sauvage}, {Savage}, {Savini},
  {Sawyer}, {Scharmberg}, {Schmitt}, {Schneider}, {Schulz}, {Schwartz},
  {Shafer}, {Shupe}, {Sibthorpe}, {Sidher}, {Smith}, {Smith}, {Smith},
  {Spencer}, {Stobie}, {Sudiwala}, {Sukhatme}, {Surace}, {Stevens}, {Swinyard},
  {Trichas}, {Tourette}, {Triou}, {Tseng}, {Tucker}, {Turner}, {Vaccari},
  {Valtchanov}, {Vigroux}, {Virique}, {Voellmer}, {Walker}, {Ward}, {Waskett},
  {Weilert}, {Wesson}, {White}, {Whitehouse}, {Wilson}, {Winter}, {Woodcraft},
  {Wright}, {Xu}, {Zavagno}, {Zemcov}, {Zhang}, \& {Zonca}}]{Griffin2010}
{Griffin}, M.~J., {Abergel}, A., {Abreu}, A., {et~al.} 2010, \aap, 518, L3+

\bibitem[{{Gruppioni} {et~al.}(2013){Gruppioni}, {Pozzi}, {Rodighiero},
  {Delvecchio}, {Berta}, {Pozzetti}, {Zamorani}, {Andreani}, {Cimatti},
  {Ilbert}, {Le Floc'h}, {Lutz}, {Magnelli}, {Marchetti}, {Monaco}, {Nordon},
  {Oliver}, {Popesso}, {Riguccini}, {Roseboom}, {Rosario}, {Sargent},
  {Vaccari}, {Altieri}, {Aussel}, {Bongiovanni}, {Cepa}, {Daddi},
  {Dom{\'{\i}}nguez-S{\'a}nchez}, {Elbaz}, {F{\"o}rster Schreiber}, {Genzel},
  {Iribarrem}, {Magliocchetti}, {Maiolino}, {Poglitsch}, {P{\'e}rez
  Garc{\'{\i}}a}, {Sanchez-Portal}, {Sturm}, {Tacconi}, {Valtchanov},
  {Amblard}, {Arumugam}, {Bethermin}, {Bock}, {Boselli}, {Buat}, {Burgarella},
  {Castro-Rodr{\'{\i}}guez}, {Cava}, {Chanial}, {Clements}, {Conley}, {Cooray},
  {Dowell}, {Dwek}, {Eales}, {Franceschini}, {Glenn}, {Griffin},
  {Hatziminaoglou}, {Ibar}, {Isaak}, {Ivison}, {Lagache}, {Levenson}, {Lu},
  {Madden}, {Maffei}, {Mainetti}, {Nguyen}, {O'Halloran}, {Page}, {Panuzzo},
  {Papageorgiou}, {Pearson}, {P{\'e}rez-Fournon}, {Pohlen}, {Rigopoulou},
  {Rowan-Robinson}, {Schulz}, {Scott}, {Seymour}, {Shupe}, {Smith}, {Stevens},
  {Symeonidis}, {Trichas}, {Tugwell}, {Vigroux}, {Wang}, {Wright}, {Xu},
  {Zemcov}, {Bardelli}, {Carollo}, {Contini}, {Le F{\'e}vre}, {Lilly},
  {Mainieri}, {Renzini}, {Scodeggio}, \& {Zucca}}]{Gruppioni2013}
{Gruppioni}, C., {Pozzi}, F., {Rodighiero}, G., {et~al.} 2013, \mnras, 432, 23

\bibitem[{{Heinis} {et~al.}(2013){Heinis}, {Buat}, {B{\'e}thermin}, {Aussel},
  {Bock}, {Boselli}, {Burgarella}, {Conley}, {Cooray}, {Farrah}, {Ibar},
  {Ilbert}, {Ivison}, {Magdis}, {Marsden}, {Oliver}, {Page}, {Rodighiero},
  {Roehlly}, {Schulz}, {Scott}, {Smith}, {Viero}, {Wang}, \&
  {Zemcov}}]{Heinis2013}
{Heinis}, S., {Buat}, V., {B{\'e}thermin}, M., {et~al.} 2013, \mnras, 429, 1113

\bibitem[{{Heinis} {et~al.}(2014){Heinis}, {Buat}, {B{\'e}thermin}, {Bock},
  {Burgarella}, {Conley}, {Cooray}, {Farrah}, {Ilbert}, {Magdis}, {Marsden},
  {Oliver}, {Rigopoulou}, {Roehlly}, {Schulz}, {Symeonidis}, {Viero}, {Xu}, \&
  {Zemcov}}]{Heinis2014}
{Heinis}, S., {Buat}, V., {B{\'e}thermin}, M., {et~al.} 2014, \mnras, 437, 1268

\bibitem[{{Ilbert} {et~al.}(2006){Ilbert}, {Arnouts}, {McCracken},
  {Bolzonella}, {Bertin}, {Le F{\`e}vre}, {Mellier}, {Zamorani}, {Pell{\`o}},
  {Iovino}, {Tresse}, {Le Brun}, {Bottini}, {Garilli}, {Maccagni}, {Picat},
  {Scaramella}, {Scodeggio}, {Vettolani}, {Zanichelli}, {Adami}, {Bardelli},
  {Cappi}, {Charlot}, {Ciliegi}, {Contini}, {Cucciati}, {Foucaud}, {Franzetti},
  {Gavignaud}, {Guzzo}, {Marano}, {Marinoni}, {Mazure}, {Meneux}, {Merighi},
  {Paltani}, {Pollo}, {Pozzetti}, {Radovich}, {Zucca}, {Bondi}, {Bongiorno},
  {Busarello}, {de La Torre}, {Gregorini}, {Lamareille}, {Mathez}, {Merluzzi},
  {Ripepi}, {Rizzo}, \& {Vergani}}]{Ilbert2006}
{Ilbert}, O., {Arnouts}, S., {McCracken}, H.~J., {et~al.} 2006, \aap, 457, 841

\bibitem[{{Ilbert} {et~al.}(2013){Ilbert}, {McCracken}, {Le F{\`e}vre},
  {Capak}, {Dunlop}, {Karim}, {Renzini}, {Caputi}, {Boissier}, {Arnouts},
  {Aussel}, {Comparat}, {Guo}, {Hudelot}, {Kartaltepe}, {Kneib}, {Krogager},
  {Le Floc'h}, {Lilly}, {Mellier}, {Milvang-Jensen}, {Moutard}, {Onodera},
  {Richard}, {Salvato}, {Sanders}, {Scoville}, {Silverman}, {Taniguchi},
  {Tasca}, {Thomas}, {Toft}, {Tresse}, {Vergani}, {Wolk}, \&
  {Zirm}}]{Ilbert2013}
{Ilbert}, O., {McCracken}, H.~J., {Le F{\`e}vre}, O., {et~al.} 2013, \aap, 556,
  A55

\bibitem[{{Ilbert} {et~al.}(2010){Ilbert}, {Salvato}, {Le Floc'h}, {Aussel},
  {Capak}, {McCracken}, {Mobasher}, {Kartaltepe}, {Scoville}, {Sanders},
  {Arnouts}, {Bundy}, {Cassata}, {Kneib}, {Koekemoer}, {Le F{\`e}vre}, {Lilly},
  {Surace}, {Taniguchi}, {Tasca}, {Thompson}, {Tresse}, {Zamojski}, {Zamorani},
  \& {Zucca}}]{Ilbert2010}
{Ilbert}, O., {Salvato}, M., {Le Floc'h}, E., {et~al.} 2010, \apj, 709, 644

\bibitem[{{Jauzac} {et~al.}(2011){Jauzac}, {Dole}, {Le Floc'h}, {Aussel},
  {Caputi}, {Ilbert}, {Salvato}, {Bavouzet}, {Beelen}, {B{\'e}thermin},
  {Kneib}, {Lagache}, \& {Puget}}]{Jauzac2011}
{Jauzac}, M., {Dole}, H., {Le Floc'h}, E., {et~al.} 2011, \aap, 525, A52+

\bibitem[{{Karim} {et~al.}(2011){Karim}, {Schinnerer},
  {Mart{\'{\i}}nez-Sansigre}, {Sargent}, {van der Wel}, {Rix}, {Ilbert},
  {Smol{\v c}i{\'c}}, {Carilli}, {Pannella}, {Koekemoer}, {Bell}, \&
  {Salvato}}]{Karim2011}
{Karim}, A., {Schinnerer}, E., {Mart{\'{\i}}nez-Sansigre}, A., {et~al.} 2011,
  \apj, 730, 61

\bibitem[{{Kennicutt}(1983)}]{Kennicutt1983}
{Kennicutt}, Jr., R.~C. 1983, \apj, 272, 54

\bibitem[{{Kennicutt}(1998)}]{Kennicutt1998}
{Kennicutt}, Jr., R.~C. 1998, \apj, 498, 541

\bibitem[{{Kewley} \& {Ellison}(2008)}]{Kewley2008}
{Kewley}, L.~J. \& {Ellison}, S.~L. 2008, \apj, 681, 1183

\bibitem[{{Kov{\'a}cs}(2006)}]{Kovacs2006}
{Kov{\'a}cs}, A. 2006, PhD thesis, Caltech

\bibitem[{{Kov{\'a}cs}(2008)}]{Kovacs2008}
{Kov{\'a}cs}, A. 2008, in Society of Photo-Optical Instrumentation Engineers
  (SPIE) Conference Series, Vol. 7020, Society of Photo-Optical Instrumentation
  Engineers (SPIE) Conference Series

\bibitem[{{Kurczynski} \& {Gawiser}(2010)}]{Kurczynski2010}
{Kurczynski}, P. \& {Gawiser}, E. 2010, \aj, 139, 1592

\bibitem[{{Lacey et al.}(2014)}]{Lacey2014}
{Lacey et al.} 2014, sub.

\bibitem[{{Lagos} {et~al.}(2012){Lagos}, {Bayet}, {Baugh}, {Lacey}, {Bell},
  {Fanidakis}, \& {Geach}}]{Lagos2012}
{Lagos}, C.~d.~P., {Bayet}, E., {Baugh}, C.~M., {et~al.} 2012, \mnras, 426,
  2142

\bibitem[{{Le Floc'h} {et~al.}(2009){Le Floc'h}, {Aussel}, {Ilbert},
  {Riguccini}, {Frayer}, {Salvato}, {Arnouts}, {Surace}, {Feruglio},
  {Rodighiero}, {Capak}, {Kartaltepe}, {Heinis}, {Sheth}, {Yan}, {McCracken},
  {Thompson}, {Sanders}, {Scoville}, \& {Koekemoer}}]{Le_Floch2009}
{Le Floc'h}, E., {Aussel}, H., {Ilbert}, O., {et~al.} 2009, \apj, 703, 222

\bibitem[{{Lee} {et~al.}(2013){Lee}, {Sanders}, {Casey}, {Scoville}, {Hung},
  {Le Floc'h}, {Ilbert}, {Aussel}, {Capak}, {Kartaltepe}, {Roseboom},
  {Salvato}, {Aravena}, {Berta}, {Bock}, {Oliver}, {Riguccini}, \&
  {Symeonidis}}]{Lee2013}
{Lee}, N., {Sanders}, D.~B., {Casey}, C.~M., {et~al.} 2013, \apj, 778, 131

\bibitem[{{Leroy} {et~al.}(2011){Leroy}, {Bolatto}, {Gordon}, {Sandstrom},
  {Gratier}, {Rosolowsky}, {Engelbracht}, {Mizuno}, {Corbelli}, {Fukui}, \&
  {Kawamura}}]{Leroy2011}
{Leroy}, A.~K., {Bolatto}, A., {Gordon}, K., {et~al.} 2011, \apj, 737, 12

\bibitem[{{Lilly} {et~al.}(2013){Lilly}, {Carollo}, {Pipino}, {Renzini}, \&
  {Peng}}]{Lilly2013}
{Lilly}, S.~J., {Carollo}, C.~M., {Pipino}, A., {Renzini}, A., \& {Peng}, Y.
  2013, \apj, 772, 119

\bibitem[{{Lutz} {et~al.}(2011){Lutz}, {Poglitsch}, {Altieri}, {Andreani},
  {Aussel}, {Berta}, {Bongiovanni}, {Brisbin}, {Cava}, {Cepa}, {Cimatti},
  {Daddi}, {Dominguez-Sanchez}, {Elbaz}, {F{\"o}rster Schreiber}, {Genzel},
  {Grazian}, {Gruppioni}, {Harwit}, {Le Floc'h}, {Magdis}, {Magnelli},
  {Maiolino}, {Nordon}, {P{\'e}rez Garc{\'{\i}}a}, {Popesso}, {Pozzi},
  {Riguccini}, {Rodighiero}, {Saintonge}, {Sanchez Portal}, {Santini}, {Shao},
  {Sturm}, {Tacconi}, {Valtchanov}, {Wetzstein}, \& {Wieprecht}}]{Lutz2011}
{Lutz}, D., {Poglitsch}, A., {Altieri}, B., {et~al.} 2011, \aap, 532, A90

\bibitem[{{Magdis} {et~al.}(2012{\natexlab{a}}){Magdis}, {Daddi},
  {B{\'e}thermin}, {Sargent}, {Elbaz}, {Pannella}, {Dickinson}, {Dannerbauer},
  {da Cunha}, {Walter}, {Rigopoulou}, {Charmandaris}, {Hwang}, \&
  {Kartaltepe}}]{Magdis2012b}
{Magdis}, G.~E., {Daddi}, E., {B{\'e}thermin}, M., {et~al.} 2012{\natexlab{a}},
  \apj, 760, 6

\bibitem[{{Magdis} {et~al.}(2012{\natexlab{b}}){Magdis}, {Daddi}, {Sargent},
  {Elbaz}, {Gobat}, {Dannerbauer}, {Feruglio}, {Tan}, {Rigopoulou},
  {Charmandaris}, {Dickinson}, {Reddy}, \& {Aussel}}]{Magdis2012a}
{Magdis}, G.~E., {Daddi}, E., {Sargent}, M., {et~al.} 2012{\natexlab{b}},
  \apjl, 758, L9

\bibitem[{{Magdis} {et~al.}(2011){Magdis}, {Elbaz}, {Dickinson}, {Hwang},
  {Charmandaris}, {Armus}, {Daddi}, {Le Floc'h}, {Aussel}, {Dannerbauer},
  {Rigopoulou}, {Buat}, {Morrison}, {Mullaney}, {Lutz}, {Scott}, {Coia},
  {Pope}, {Pannella}, {Altieri}, {Burgarella}, {Bethermin}, {Dasyra},
  {Kartaltepe}, {Leiton}, {Magnelli}, {Popesso}, \& {Valtchanov}}]{Magdis2011}
{Magdis}, G.~E., {Elbaz}, D., {Dickinson}, M., {et~al.} 2011, \aap, 534, A15

\bibitem[{{Magdis} {et~al.}(2010){Magdis}, {Rigopoulou}, {Huang}, \&
  {Fazio}}]{Magdis2010}
{Magdis}, G.~E., {Rigopoulou}, D., {Huang}, J.-S., \& {Fazio}, G.~G. 2010,
  \mnras, 401, 1521

\bibitem[{{Magnelli} {et~al.}(2013){Magnelli}, {Popesso}, {Berta}, {Pozzi},
  {Elbaz}, {Lutz}, {Dickinson}, {Altieri}, {Andreani}, {Aussel},
  {B{\'e}thermin}, {Bongiovanni}, {Cepa}, {Charmandaris}, {Chary}, {Cimatti},
  {Daddi}, {F{\"o}rster Schreiber}, {Genzel}, {Gruppioni}, {Harwit}, {Hwang},
  {Ivison}, {Magdis}, {Maiolino}, {Murphy}, {Nordon}, {Pannella}, {P{\'e}rez
  Garc{\'{\i}}a}, {Poglitsch}, {Rosario}, {Sanchez-Portal}, {Santini}, {Scott},
  {Sturm}, {Tacconi}, \& {Valtchanov}}]{Magnelli2013}
{Magnelli}, B., {Popesso}, P., {Berta}, S., {et~al.} 2013, \aap, 553, A132

\bibitem[{{Mannucci} {et~al.}(2010){Mannucci}, {Cresci}, {Maiolino}, {Marconi},
  \& {Gnerucci}}]{Mannucci2010}
{Mannucci}, F., {Cresci}, G., {Maiolino}, R., {Marconi}, A., \& {Gnerucci}, A.
  2010, \mnras, 408, 2115

\bibitem[{{McCracken} {et~al.}(2012){McCracken}, {Milvang-Jensen}, {Dunlop},
  {Franx}, {Fynbo}, {Le F{\`e}vre}, {Holt}, {Caputi}, {Goranova}, {Buitrago},
  {Emerson}, {Freudling}, {Hudelot}, {L{\'o}pez-Sanjuan}, {Magnard}, {Mellier},
  {M{\o}ller}, {Nilsson}, {Sutherland}, {Tasca}, \& {Zabl}}]{McCracken2012}
{McCracken}, H.~J., {Milvang-Jensen}, B., {Dunlop}, J., {et~al.} 2012, \aap,
  544, A156

\bibitem[{{Mitchell} {et~al.}(2013){Mitchell}, {Lacey}, {Baugh}, \&
  {Cole}}]{Mitchell2013}
{Mitchell}, P.~D., {Lacey}, C.~G., {Baugh}, C.~M., \& {Cole}, S. 2013, \mnras,
  435, 87

\bibitem[{{M{\o}ller} {et~al.}(2013){M{\o}ller}, {Fynbo}, {Ledoux}, \&
  {Nilsson}}]{Moller2013}
{M{\o}ller}, P., {Fynbo}, J.~P.~U., {Ledoux}, C., \& {Nilsson}, K.~K. 2013,
  \mnras, 430, 2680

\bibitem[{{Moster} {et~al.}(2013){Moster}, {Naab}, \& {White}}]{Moster2013}
{Moster}, B.~P., {Naab}, T., \& {White}, S.~D.~M. 2013, \mnras, 428, 3121

\bibitem[{{Moster} {et~al.}(2010){Moster}, {Somerville}, {Maulbetsch}, {van den
  Bosch}, {Macci{\`o}}, {Naab}, \& {Oser}}]{Moster2010}
{Moster}, B.~P., {Somerville}, R.~S., {Maulbetsch}, C., {et~al.} 2010, \apj,
  710, 903

\bibitem[{{Mu{\~n}oz-Mateos} {et~al.}(2009){Mu{\~n}oz-Mateos}, {Gil de Paz},
  {Boissier}, {Zamorano}, {Dale}, {P{\'e}rez-Gonz{\'a}lez}, {Gallego},
  {Madore}, {Bendo}, {Thornley}, {Draine}, {Boselli}, {Buat}, {Calzetti},
  {Moustakas}, \& {Kennicutt}}]{Munoz-Mateos2009}
{Mu{\~n}oz-Mateos}, J.~C., {Gil de Paz}, A., {Boissier}, S., {et~al.} 2009,
  \apj, 701, 1965

\bibitem[{{Nagao} {et~al.}(2011){Nagao}, {Maiolino}, {Marconi}, \&
  {Matsuhara}}]{Nagao2011}
{Nagao}, T., {Maiolino}, R., {Marconi}, A., \& {Matsuhara}, H. 2011, \aap, 526,
  A149

\bibitem[{{Nguyen} {et~al.}(2010){Nguyen}, {Schulz}, {Levenson}, {Amblard},
  {Arumugam}, {Aussel}, {Babbedge}, {Blain}, {Bock}, {Boselli}, {Buat},
  {Castro-Rodriguez}, {Cava}, {Chanial}, {Chapin}, {Clements}, {Conley},
  {Conversi}, {Cooray}, {Dowell}, {Dwek}, {Eales}, {Elbaz}, {Fox},
  {Franceschini}, {Gear}, {Glenn}, {Griffin}, {Halpern}, {Hatziminaoglou},
  {Ibar}, {Isaak}, {Ivison}, {Lagache}, {Lu}, {Madden}, {Maffei}, {Mainetti},
  {Marchetti}, {Marsden}, {Marshall}, {O'Halloran}, {Oliver}, {Omont}, {Page},
  {Panuzzo}, {Papageorgiou}, {Pearson}, {Perez Fournon}, {Pohlen}, {Rangwala},
  {Rigopoulou}, {Rizzo}, {Roseboom}, {Rowan-Robinson}, {Scott}, {Seymour},
  {Shupe}, {Smith}, {Stevens}, {Symeonidis}, {Trichas}, {Tugwell}, {Vaccari},
  {Valtchanov}, {Vigroux}, {Wang}, {Ward}, {Wiebe}, {Wright}, {Xu}, \&
  {Zemcov}}]{Nguyen2010}
{Nguyen}, H.~T., {Schulz}, B., {Levenson}, L., {et~al.} 2010, \aap, 518, L5

\bibitem[{{Noeske} {et~al.}(2007){Noeske}, {Weiner}, {Faber}, {Papovich},
  {Koo}, {Somerville}, {Bundy}, {Conselice}, {Newman}, {Schiminovich}, {Le
  Floc'h}, {Coil}, {Rieke}, {Lotz}, {Primack}, {Barmby}, {Cooper}, {Davis},
  {Ellis}, {Fazio}, {Guhathakurta}, {Huang}, {Kassin}, {Martin}, {Phillips},
  {Rich}, {Small}, {Willmer}, \& {Wilson}}]{Noeske2007}
{Noeske}, K.~G., {Weiner}, B.~J., {Faber}, S.~M., {et~al.} 2007, \apjl, 660,
  L43

\bibitem[{{Obreschkow} \& {Rawlings}(2009)}]{Obreschkow2009}
{Obreschkow}, D. \& {Rawlings}, S. 2009, \apjl, 696, L129

\bibitem[{{Oliver} {et~al.}(2012){Oliver}, {Bock}, {Altieri}, {Amblard},
  {Arumugam}, {Aussel}, {Babbedge}, {Beelen}, {B{\'e}thermin}, {Blain},
  {Boselli}, {Bridge}, {Brisbin}, {Buat}, {Burgarella},
  {Castro-Rodr{\'{\i}}guez}, {Cava}, {Chanial}, {Cirasuolo}, {Clements},
  {Conley}, {Conversi}, {Cooray}, {Dowell}, {Dubois}, {Dwek}, {Dye}, {Eales},
  {Elbaz}, {Farrah}, {Feltre}, {Ferrero}, {Fiolet}, {Fox}, {Franceschini},
  {Gear}, {Giovannoli}, {Glenn}, {Gong}, {Gonz{\'a}lez Solares}, {Griffin},
  {Halpern}, {Harwit}, {Hatziminaoglou}, {Heinis}, {Hurley}, {Hwang}, {Hyde},
  {Ibar}, {Ilbert}, {Isaak}, {Ivison}, {Lagache}, {Le Floc'h}, {Levenson},
  {Faro}, {Lu}, {Madden}, {Maffei}, {Magdis}, {Mainetti}, {Marchetti},
  {Marsden}, {Marshall}, {Mortier}, {Nguyen}, {O'Halloran}, {Omont}, {Page},
  {Panuzzo}, {Papageorgiou}, {Patel}, {Pearson}, {P{\'e}rez-Fournon}, {Pohlen},
  {Rawlings}, {Raymond}, {Rigopoulou}, {Riguccini}, {Rizzo}, {Rodighiero},
  {Roseboom}, {Rowan-Robinson}, {S{\'a}nchez Portal}, {Schulz}, {Scott},
  {Seymour}, {Shupe}, {Smith}, {Stevens}, {Symeonidis}, {Trichas}, {Tugwell},
  {Vaccari}, {Valtchanov}, {Vieira}, {Viero}, {Vigroux}, {Wang}, {Ward},
  {Wardlow}, {Wright}, {Xu}, \& {Zemcov}}]{Oliver2012}
{Oliver}, S.~J., {Bock}, J., {Altieri}, B., {et~al.} 2012, \mnras, 424, 1614

\bibitem[{{Pannella} {et~al.}(2009){Pannella}, {Carilli}, {Daddi}, {McCracken},
  {Owen}, {Renzini}, {Strazzullo}, {Civano}, {Koekemoer}, {Schinnerer},
  {Scoville}, {Smol{\v c}i{\'c}}, {Taniguchi}, {Aussel}, {Kneib}, {Ilbert},
  {Mellier}, {Salvato}, {Thompson}, \& {Willott}}]{Pannella2009}
{Pannella}, M., {Carilli}, C.~L., {Daddi}, E., {et~al.} 2009, \apjl, 698, L116

\bibitem[{{Pannella} {et~al.}(2014){Pannella}, {Elbaz}, {Daddi}, {Dickinson},
  {Hwang}, {Schreiber}, {Strazzullo}, {Aussel}, {Bethermin}, {Buat},
  {Charmandaris}, {Cibinel}, {Juneau}, {Ivison}, {Le Borgne}, {Le Floc'h},
  {Leiton}, {Lin}, {Magdis}, {Morrison}, {Mullaney}, {Onodera}, {Renzini},
  {Salim}, {Sargent}, {Scott}, {Shu}, \& {Wang}}]{Pannella2014}
{Pannella}, M., {Elbaz}, D., {Daddi}, E., {et~al.} 2014, arXiv:1407.5072

\bibitem[{{Peng} {et~al.}(2010){Peng}, {Lilly}, {Kova{\v c}}, {Bolzonella},
  {Pozzetti}, {Renzini}, {Zamorani}, {Ilbert}, {Knobel}, {Iovino}, {Maier},
  {Cucciati}, {Tasca}, {Carollo}, {Silverman}, {Kampczyk}, {de Ravel},
  {Sanders}, {Scoville}, {Contini}, {Mainieri}, {Scodeggio}, {Kneib}, {Le
  F{\`e}vre}, {Bardelli}, {Bongiorno}, {Caputi}, {Coppa}, {de la Torre},
  {Franzetti}, {Garilli}, {Lamareille}, {Le Borgne}, {Le Brun}, {Mignoli},
  {Perez Montero}, {Pello}, {Ricciardelli}, {Tanaka}, {Tresse}, {Vergani},
  {Welikala}, {Zucca}, {Oesch}, {Abbas}, {Barnes}, {Bordoloi}, {Bottini},
  {Cappi}, {Cassata}, {Cimatti}, {Fumana}, {Hasinger}, {Koekemoer},
  {Leauthaud}, {Maccagni}, {Marinoni}, {McCracken}, {Memeo}, {Meneux}, {Nair},
  {Porciani}, {Presotto}, \& {Scaramella}}]{Peng2010}
{Peng}, Y.-j., {Lilly}, S.~J., {Kova{\v c}}, K., {et~al.} 2010, \apj, 721, 193

\bibitem[{{Pilbratt} {et~al.}(2010){Pilbratt}, {Riedinger}, {Passvogel},
  {Crone}, {Doyle}, {Gageur}, {Heras}, {Jewell}, {Metcalfe}, {Ott}, \&
  {Schmidt}}]{Pilbratt2010}
{Pilbratt}, G.~L., {Riedinger}, J.~R., {Passvogel}, T., {et~al.} 2010, \aap,
  518, L1+

\bibitem[{{Planck Collaboration} {et~al.}(2013){Planck Collaboration}, {Ade},
  {Aghanim}, {Armitage-Caplan}, {Arnaud}, {Ashdown}, {Atrio-Barandela},
  {Aumont}, {Baccigalupi}, {Banday}, \& et~al.}]{Planck_CIB_2013}
{Planck Collaboration}, {Ade}, P.~A.~R., {Aghanim}, N., {et~al.} 2013,
  arXiv:1309.0382

\bibitem[{{Poglitsch} {et~al.}(2010){Poglitsch}, {Waelkens}, {Geis},
  {Feuchtgruber}, {Vandenbussche}, {Rodriguez}, {Krause}, {Renotte}, {van
  Hoof}, {Saraceno}, {Cepa}, {Kerschbaum}, {Agn{\`e}se}, {Ali}, {Altieri},
  {Andreani}, {Augueres}, {Balog}, {Barl}, {Bauer}, {Belbachir}, {Benedettini},
  {Billot}, {Boulade}, {Bischof}, {Blommaert}, {Callut}, {Cara}, {Cerulli},
  {Cesarsky}, {Contursi}, {Creten}, {De Meester}, {Doublier}, {Doumayrou},
  {Duband}, {Exter}, {Genzel}, {Gillis}, {Gr{\"o}zinger}, {Henning},
  {Herreros}, {Huygen}, {Inguscio}, {Jakob}, {Jamar}, {Jean}, {de Jong},
  {Katterloher}, {Kiss}, {Klaas}, {Lemke}, {Lutz}, {Madden}, {Marquet},
  {Martignac}, {Mazy}, {Merken}, {Montfort}, {Morbidelli}, {M{\"u}ller},
  {Nielbock}, {Okumura}, {Orfei}, {Ottensamer}, {Pezzuto}, {Popesso},
  {Putzeys}, {Regibo}, {Reveret}, {Royer}, {Sauvage}, {Schreiber}, {Stegmaier},
  {Schmitt}, {Schubert}, {Sturm}, {Thiel}, {Tofani}, {Vavrek}, {Wetzstein},
  {Wieprecht}, \& {Wiezorrek}}]{Poglitsch2010}
{Poglitsch}, A., {Waelkens}, C., {Geis}, N., {et~al.} 2010, \aap, 518, L2+

\bibitem[{{Popesso} {et~al.}(2012){Popesso}, {Magnelli}, {Buttiglione}, {Lutz},
  {Poglitsch}, {Berta}, {Nordon}, {Altieri}, {Aussel}, {Billot}, {Gastaud},
  {Ali}, {Balog}, {Cava}, {Feuchtgruber}, {Gonzalez Garcia}, {Geis}, {Kiss},
  {Klaas}, {Linz}, {Liu}, {Moor}, {Morin}, {Muller}, {Nielbock}, {Okumura},
  {Osterhage}, {Ottensamer}, {Paladini}, {Pezzuto}, {Dublier Pritchard},
  {Regibo}, {Rodighiero}, {Royer}, {Sauvage}, {Sturm}, {Wetzstein},
  {Wieprecht}, \& {Wiezorrek}}]{Popesso2012}
{Popesso}, P., {Magnelli}, B., {Buttiglione}, S., {et~al.} 2012,
  arXiv:1211.4257

\bibitem[{{R{\'e}my-Ruyer} {et~al.}(2014){R{\'e}my-Ruyer}, {Madden},
  {Galliano}, {Galametz}, {Takeuchi}, {Asano}, {Zhukovska}, {Lebouteiller},
  {Cormier}, {Jones}, {Bocchio}, {Baes}, {Bendo}, {Boquien}, {Boselli},
  {DeLooze}, {Doublier-Pritchard}, {Hughes}, {Karczewski}, \&
  {Spinoglio}}]{Remy2014}
{R{\'e}my-Ruyer}, A., {Madden}, S.~C., {Galliano}, F., {et~al.} 2014, \aap,
  563, A31

\bibitem[{{Riechers} {et~al.}(2013){Riechers}, {Bradford}, {Clements},
  {Dowell}, {P{\'e}rez-Fournon}, {Ivison}, {Bridge}, {Conley}, {Fu}, {Vieira},
  {Wardlow}, {Calanog}, {Cooray}, {Hurley}, {Neri}, {Kamenetzky}, {Aguirre},
  {Altieri}, {Arumugam}, {Benford}, {B{\'e}thermin}, {Bock}, {Burgarella},
  {Cabrera-Lavers}, {Chapman}, {Cox}, {Dunlop}, {Earle}, {Farrah}, {Ferrero},
  {Franceschini}, {Gavazzi}, {Glenn}, {Solares}, {Gurwell}, {Halpern},
  {Hatziminaoglou}, {Hyde}, {Ibar}, {Kov{\'a}cs}, {Krips}, {Lupu}, {Maloney},
  {Martinez-Navajas}, {Matsuhara}, {Murphy}, {Naylor}, {Nguyen}, {Oliver},
  {Omont}, {Page}, {Petitpas}, {Rangwala}, {Roseboom}, {Scott}, {Smith},
  {Staguhn}, {Streblyanska}, {Thomson}, {Valtchanov}, {Viero}, {Wang},
  {Zemcov}, \& {Zmuidzinas}}]{Riechers2013}
{Riechers}, D.~A., {Bradford}, C.~M., {Clements}, D.~L., {et~al.} 2013, \nat,
  496, 329

\bibitem[{{Rodighiero} {et~al.}(2011){Rodighiero}, {Daddi}, {Baronchelli},
  {Cimatti}, {Renzini}, {Aussel}, {Popesso}, {Lutz}, {Andreani}, {Berta},
  {Cava}, {Elbaz}, {Feltre}, {Fontana}, {F{\"o}rster Schreiber},
  {Franceschini}, {Genzel}, {Grazian}, {Gruppioni}, {Ilbert}, {Le Floch},
  {Magdis}, {Magliocchetti}, {Magnelli}, {Maiolino}, {McCracken}, {Nordon},
  {Poglitsch}, {Santini}, {Pozzi}, {Riguccini}, {Tacconi}, {Wuyts}, \&
  {Zamorani}}]{Rodighiero2011}
{Rodighiero}, G., {Daddi}, E., {Baronchelli}, I., {et~al.} 2011, \apjl, 739,
  L40

\bibitem[{{Saintonge} {et~al.}(2013){Saintonge}, {Lutz}, {Genzel}, {Magnelli},
  {Nordon}, {Tacconi}, {Baker}, {Bandara}, {Berta}, {F{\"o}rster Schreiber},
  {Poglitsch}, {Sturm}, {Wuyts}, \& {Wuyts}}]{Saintonge2013}
{Saintonge}, A., {Lutz}, D., {Genzel}, R., {et~al.} 2013, \apj, 778, 2

\bibitem[{{Santini} {et~al.}(2014){Santini}, {Maiolino}, {Magnelli}, {Lutz},
  {Lamastra}, {Li Causi}, {Eales}, {Andreani}, {Berta}, {Buat}, {Cooray},
  {Cresci}, {Daddi}, {Farrah}, {Fontana}, {Franceschini}, {Genzel}, {Granato},
  {Grazian}, {Le Floc'h}, {Magdis}, {Magliocchetti}, {Mannucci}, {Menci},
  {Nordon}, {Oliver}, {Popesso}, {Pozzi}, {Riguccini}, {Rodighiero}, {Rosario},
  {Salvato}, {Scott}, {Silva}, {Tacconi}, {Viero}, {Wang}, {Wuyts}, \&
  {Xu}}]{Santini2014}
{Santini}, P., {Maiolino}, R., {Magnelli}, B., {et~al.} 2014, \aap, 562, A30

\bibitem[{{Sargent} {et~al.}(2012){Sargent}, {B{\'e}thermin}, {Daddi}, \&
  {Elbaz}}]{Sargent2012}
{Sargent}, M.~T., {B{\'e}thermin}, M., {Daddi}, E., \& {Elbaz}, D. 2012, \apjl,
  747, L31

\bibitem[{{Sargent} {et~al.}(2014){Sargent}, {Daddi}, {B{\'e}thermin},
  {Aussel}, {Magdis}, {Hwang}, {Juneau}, {Elbaz}, \& {da Cunha}}]{Sargent2014}
{Sargent}, M.~T., {Daddi}, E., {B{\'e}thermin}, M., {et~al.} 2014, \apj, 793,
  19

\bibitem[{{Scoville} {et~al.}(2014){Scoville}, {Aussel}, {Sheth}, {Scott},
  {Sanders}, {Ivison}, {Pope}, {Capak}, {Vanden Bout}, {Manohar}, {Kartaltepe},
  {Robertson}, \& {Lilly}}]{Scoville2014}
{Scoville}, N., {Aussel}, H., {Sheth}, K., {et~al.} 2014, \apj, 783, 84

\bibitem[{{Steidel} {et~al.}(2014){Steidel}, {Rudie}, {Strom}, {Pettini},
  {Reddy}, {Shapley}, {Trainor}, {Erb}, {Turner}, {Konidaris}, {Kulas}, {Mace},
  {Matthews}, \& {McLean}}]{Steidel2014}
{Steidel}, C.~C., {Rudie}, G.~C., {Strom}, A.~L., {et~al.} 2014,
  arXiv:1405.5473

\bibitem[{{Symeonidis} {et~al.}(2013){Symeonidis}, {Vaccari}, {Berta}, {Page},
  {Lutz}, {Arumugam}, {Aussel}, {Bock}, {Boselli}, {Buat}, {Capak}, {Clements},
  {Conley}, {Conversi}, {Cooray}, {Dowell}, {Farrah}, {Franceschini},
  {Giovannoli}, {Glenn}, {Griffin}, {Hatziminaoglou}, {Hwang}, {Ibar},
  {Ilbert}, {Ivison}, {Floc'h}, {Lilly}, {Kartaltepe}, {Magnelli}, {Magdis},
  {Marchetti}, {Nguyen}, {Nordon}, {O'Halloran}, {Oliver}, {Omont},
  {Papageorgiou}, {Patel}, {Pearson}, {P{\'e}rez-Fournon}, {Pohlen}, {Popesso},
  {Pozzi}, {Rigopoulou}, {Riguccini}, {Rosario}, {Roseboom}, {Rowan-Robinson},
  {Salvato}, {Schulz}, {Scott}, {Seymour}, {Shupe}, {Smith}, {Valtchanov},
  {Wang}, {Xu}, {Zemcov}, \& {Wuyts}}]{Symeonidis2013}
{Symeonidis}, M., {Vaccari}, M., {Berta}, S., {et~al.} 2013, \mnras, 431, 2317

\bibitem[{{Tacconi} {et~al.}(2010){Tacconi}, {Genzel}, {Neri}, {Cox}, {Cooper},
  {Shapiro}, {Bolatto}, {Bouch{\'e}}, {Bournaud}, {Burkert}, {Combes},
  {Comerford}, {Davis}, {Schreiber}, {Garcia-Burillo}, {Gracia-Carpio}, {Lutz},
  {Naab}, {Omont}, {Shapley}, {Sternberg}, \& {Weiner}}]{Tacconi2010}
{Tacconi}, L.~J., {Genzel}, R., {Neri}, R., {et~al.} 2010, \nat, 463, 781

\bibitem[{{Tacconi} {et~al.}(2013){Tacconi}, {Neri}, {Genzel}, {Combes},
  {Bolatto}, {Cooper}, {Wuyts}, {Bournaud}, {Burkert}, {Comerford}, {Cox},
  {Davis}, {F{\"o}rster Schreiber}, {Garc{\'{\i}}a-Burillo}, {Gracia-Carpio},
  {Lutz}, {Naab}, {Newman}, {Omont}, {Saintonge}, {Shapiro Griffin}, {Shapley},
  {Sternberg}, \& {Weiner}}]{Tacconi2013}
{Tacconi}, L.~J., {Neri}, R., {Genzel}, R., {et~al.} 2013, \apj, 768, 74

\bibitem[{{Tan} {et~al.}(2014){Tan}, {Daddi}, {Magdis}, {Pannella}, {Sargent},
  {Riechers}, {B{\'e}thermin}, {Bournaud}, {Carilli}, {da Cunha},
  {Dannerbauer}, {Dickinson}, {Elbaz}, {Gao}, {Hodge}, {Owen}, \&
  {Walter}}]{Tan2014}
{Tan}, Q., {Daddi}, E., {Magdis}, G., {et~al.} 2014, arXiv:1403.7992

\bibitem[{{Tan} {et~al.}(2013){Tan}, {Daddi}, {Sargent}, {Magdis}, {Hodge},
  {B{\'e}thermin}, {Bournaud}, {Carilli}, {Dannerbauer}, {Dickinson}, {Elbaz},
  {Gao}, {Morrison}, {Owen}, {Pannella}, {Riechers}, \& {Walter}}]{Tan2013}
{Tan}, Q., {Daddi}, E., {Sargent}, M., {et~al.} 2013, \apjl, 776, L24

\bibitem[{{Toft} {et~al.}(2014){Toft}, {Smol{\v c}i{\'c}}, {Magnelli}, {Karim},
  {Zirm}, {Michalowski}, {Capak}, {Sheth}, {Schawinski}, {Krogager}, {Wuyts},
  {Sanders}, {Man}, {Lutz}, {Staguhn}, {Berta}, {Mccracken}, {Krpan}, \&
  {Riechers}}]{Toft2014}
{Toft}, S., {Smol{\v c}i{\'c}}, V., {Magnelli}, B., {et~al.} 2014, \apj, 782,
  68

\bibitem[{{Troncoso} {et~al.}(2014){Troncoso}, {Maiolino}, {Sommariva},
  {Cresci}, {Mannucci}, {Marconi}, {Meneghetti}, {Grazian}, {Cimatti},
  {Fontana}, {Nagao}, \& {Pentericci}}]{Troncoso2014}
{Troncoso}, P., {Maiolino}, R., {Sommariva}, V., {et~al.} 2014, \aap, 563, A58

\bibitem[{{Viero} {et~al.}(2013){Viero}, {Moncelsi}, {Quadri}, {Arumugam},
  {Assef}, {B{\'e}thermin}, {Bock}, {Bridge}, {Casey}, {Conley}, {Cooray},
  {Farrah}, {Glenn}, {Heinis}, {Ibar}, {Ikarashi}, {Ivison}, {Kohno},
  {Marsden}, {Oliver}, {Roseboom}, {Schulz}, {Scott}, {Serra}, {Vaccari},
  {Vieira}, {Wang}, {Wardlow}, {Wilson}, {Yun}, \& {Zemcov}}]{Viero2013b}
{Viero}, M.~P., {Moncelsi}, L., {Quadri}, R.~F., {et~al.} 2013, \apj, 779, 32

\bibitem[{{Wang} {et~al.}(2013){Wang}, {Farrah}, {Oliver}, {Amblard},
  {B{\'e}thermin}, {Bock}, {Conley}, {Cooray}, {Halpern}, {Heinis}, {Ibar},
  {Ilbert}, {Ivison}, {Marsden}, {Roseboom}, {Rowan-Robinson}, {Schulz},
  {Smith}, {Viero}, \& {Zemcov}}]{Wang2013}
{Wang}, L., {Farrah}, D., {Oliver}, S.~J., {et~al.} 2013, \mnras, 431, 648

\bibitem[{{Welikala} {et~al.}(2014){Welikala}, {Bethermin}, {Guery},
  {Strandet}, {Aravena}, {Ashby}, \& {al.}}]{Welikala2014}
{Welikala}, N., {Bethermin}, M., {Guery}, D., {et~al.} 2014, sub. to \mnras

\bibitem[{{Whitaker} {et~al.}(2012){Whitaker}, {van Dokkum}, {Brammer}, \&
  {Franx}}]{Whitaker2012}
{Whitaker}, K.~E., {van Dokkum}, P.~G., {Brammer}, G., \& {Franx}, M. 2012,
  \apjl, 754, L29

\bibitem[{{Williams} {et~al.}(2009){Williams}, {Quadri}, {Franx}, {van Dokkum},
  \& {Labb{\'e}}}]{Williams2009}
{Williams}, R.~J., {Quadri}, R.~F., {Franx}, M., {van Dokkum}, P., \&
  {Labb{\'e}}, I. 2009, \apj, 691, 1879

\end{thebibliography}

\begin{appendix}

\section{Estimation and correction on the bias caused by the galaxy clustering on the stacking results}

\label{Annexestacking}

As explained in Sect.\,\ref{sect:stacking}, the standard stacking technique can be strongly affected by the bias caused by the clustering of the galaxies. We use two independent methods to estimate and correct it. 

\subsection{Estimation of the bias using a simulation based on the real catalog}

\label{sect:simu}

We performed an estimate of the bias induced by the clustering using a realistic simulation of the COSMOS field based on the positions and stellar masses of the real sources. The flux of each source in this simulation is estimated using the ratio between the mean far-IR/(sub-)mm fluxes and the stellar mass found by a first stacking analysis. The galaxies classified as passive are not taken into account in this simulation. This technique assumes implicitly a flat sSFR-M$_\star$ relation, since we use a constant SFR/M$_\star$ ratio versus stellar mass at fixed redshift. However, we checked that using a more standard sSFR$\propto$M$_\star^{-0.2}$ relation \citep[e.g.,][]{Rodighiero2011} has a negligible impact on the results. We applied no scatter around this relation in our simulation for simplicity. As mean stacking is a linear operation, the presence or not of a scatter has no impact on the results \citep{Bethermin2012b}.\\

A simulated map is thus produced using all the star-forming galaxies of the \citet{Ilbert2013} catalog. In order to avoid edge effects (absence of sources and thus a lower background caused by the faint unresolved sources in the region covered by the optical/near-IR data), we fill the uncovered regions drawing with replacement sources from the UltraVISTA field and putting them at a random position. The number of drawn sources is chosen to have exactly the same number density inside and outside the UltraVISTA field.\\

\begin{figure}
\centering
\includegraphics{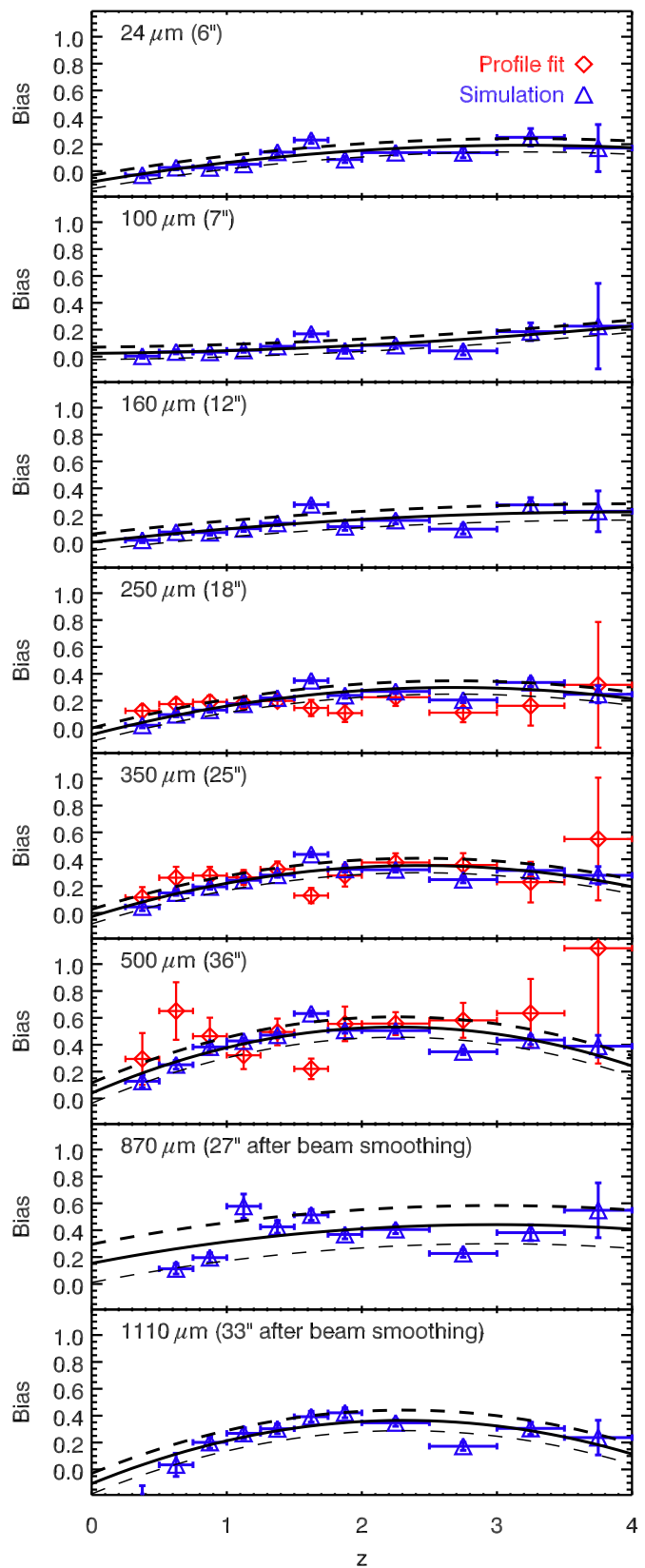}
\caption{\label{fig:clusbias} Relative bias induced by the clustering as a function of redshift at the various wavelengths we used in our analysis. The FWHM of the beam is provided in brackets. The blue triangles are the estimations from the simulation (Sect.\,\ref{sect:simu}) and the red diamonds are provided by the fit of the clustering component in map space (Sect.\,\ref{sect:fitclus}). These numbers are only valid for a complete sample of M$_\star > 3 \times 10^{10}$\,M$_\odot$ galaxies.}
\end{figure}

Finally, we measured the mean fluxes of the M$_\star>$3$\times$10$^{10}$\,M$_\odot$ sources by stacking in the simulated maps, using exactly the same photometric method as for the real data. We finally computed the relative bias between the recovered flux and the input flux ($S_{\rm out}/S_{\rm in}-1$). The results are shown Fig.\,\ref{fig:clusbias} (blue triangles). The uncertainties are computed a bootstrap method. As expected, the bias increases with the size of the beam. We can see a rise of the bias with redshift up to z$\sim$2. This trend can be understood considering the rise of the clustering of the galaxy responsible for the cosmic infrared background \citep{Planck_CIB_2013} and a rather stable number density of emitters especially below z=1 \citep{Bethermin2011,Magnelli2013,Gruppioni2013}. At higher redshift, we found a slow decrease. This trend is probably driven by the decrease in the infrared luminosity density at high redshift \citep{Planck_CIB_2013,Burgarella2013} combined with the decrease in the number density of infrared emitters \citep{Gruppioni2013}.\\ 

\subsection{Estimation of the bias fitting the clustering contribution in the stacked images}

\label{sect:fitclus}

The method presented in the previous section only takes into account the contamination of the stacks by known sources. However, faint galaxy populations could have a non negligible contribution, despite their total contribution to the infrared luminosity and their clustering are expected to be small. We thus used a second method to estimate the bias caused by the clustering which takes into account a potential contamination by these low-mass galaxies. This method is based on a simultaneous fit in the stacked images of three components: a point source at the center of the image, a clustering contamination, and a background. This technique was already successfully used by several previous works based on \textit{Herschel} and \textit{Planck} data \citep{Bethermin2012b,Heinis2013,Heinis2014,Welikala2014}.\\

In presence of clustering, the outcome of a stacking is not only a PSF with the mean flux of the population and a constant background (corresponding to the surface brightness of all galaxy populations i.e., the cosmic infrared background). There is in addition a signal coming from the greater probability of finding another neighboring infrared galaxy compared to the field because of galaxy clustering. The signal in the stacked image can thus be modeled by \citep{Bavouzet2008,Bethermin2010b}
\begin{equation}
m(x,y) = \alpha \times \textrm{PSF}(x,y) + \beta \times (\textrm{PSF} \ast w)(x,y) +\gamma,
\end{equation} 
where $m$ is the stacked image, PSF the point spread function, and $w$ the auto-correlation function. The symbol $\ast$ represents the convolution. $\alpha$, $\beta$, and $\gamma$ are free parameters corresponding to the intensity of the mean flux of the population, the clustering signal, and the background, respectively. This method works only if the PSF is well-known, the extension of the sources is negligible compared to the PSF, and the effects of the filtering are small at the scale of the stacked image. Consequently, we applied this method only to the SPIRE data for which these hypotheses are the most solid. The uncertainties on the clustering bias ($\beta / \alpha$ for the photometry we chose to use for SPIRE data) are estimated fitting the model described previously on a set of stacked images produced from 1000 bootstrap samples. The results are shown in Fig.\,\ref{fig:clusbias} (red diamonds).\\  

\subsection{Corrections of the measurements}

In Fig.\,\ref{fig:clusbias}, we can see that the two methods provide globally consistent estimates. This confirms that the low-mass galaxies not included in the UltraVISTA catalog have a minor impact. We found few outliers for which the two methods disagree. In particular, in the 1.5$<$z$<$1.75 bin, the estimation from the simulation is higher than the trend of the redshift evolution at all wavelengths, and the results from the profile fitting are lower. This could be caused, as instance, by a structures in the field or a systematic effect in the photometric redshift. Because of these few catastrophic outliers, we chose to use a correction computed from a fit of the redshift evolution of the bias instead of an individual estimate in each redshift slice.\\ 

The evolution of the bias with redshift is fitted independently at each wavelength. We chose to use a simple, second-order, polynomial model ($a z^2 + bz + c$). We used only the results from the simulation to have a consistent treatment of the various wavelengths. The scatter of the residuals is larger than the residuals, probably because bootstrap does not take into account the variance coming from the large-scale structures. We thus used the scatter of the residuals to obtain a conservative estimate of the uncertainties on the bias. In Fig.\,\ref{fig:clusbias}, the best fit is represented by a solid line and the 1$\sigma$ confidence region by a dashed line.\\

In a few case, the bias at z$>$3 can converge to unphysical negative values. We then apply no corrections, but combine the typical uncertainty on the bias to the error bars. A special treatment is also applied to the samples of strong starbursts. Their flux is typically 10 times brighter in infrared by construction (their sSFR is 10 times larger than the main sequence). In contrast, the clustering signal is not expected to be significantly stronger, because the clustering of massive starbursts and main-sequence galaxies is relatively similar \citep{Bethermin2014}. We thus divide the bias found for the full population of galaxy by a factor of 10 to estimate the one of the starbursts for simplicity.\\

\subsection{Testing another method}

We also tried to apply the \textsc{simstack} algorithm \citep{Viero2013b} to our data. This algorithm is adapted from \citet{Kurczynski2010} and uses the position of the known sources to deblend their contamination. Contrary to \citet{Kurczynski2010}, \textsc{simstack} can consider a large set of distinct galaxy populations. The mean flux of the each population is used to estimate how sources contaminate their neighbors. All populations are treated simultaneously. This is the equivalent of PSF-fitting codes but applied to a full population instead of each source individually. Unfortunately, this method is not totally unbiased in our case. We found biases up to 15\% running \textsc{simstack} on the simulation presented in Sect.\,\ref{sect:simu}, probably because the catalog of mass-selected sources is not available around bright sources. At the edge of the optical/near-IR-covered region, the flux coming from the sources outside the covered area is not corrected, when the flux from all neighbors is taken into account at the middle of zone where the mass catalog is extracted. Indeed, the algorithm works correctly if we put on the simulation only sources present in the input catalog.\\

\section{Fit residuals}

\label{sect:residuals}

\begin{figure*}
\centering
\includegraphics{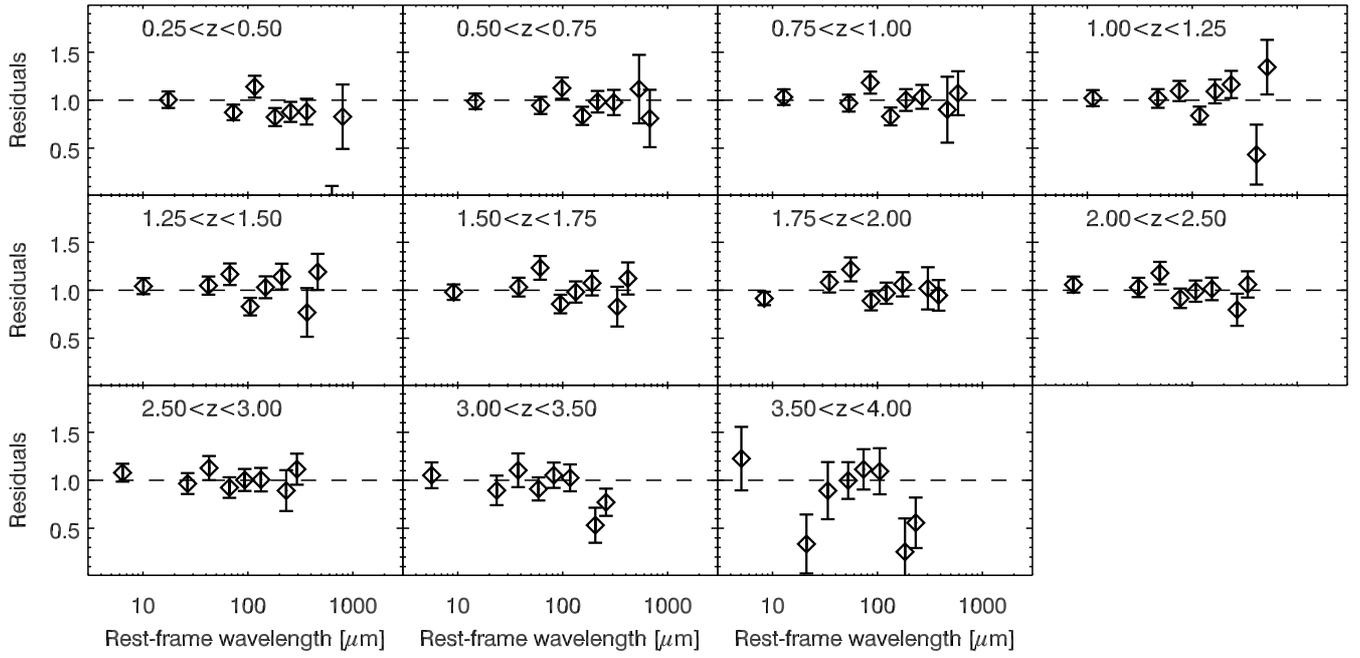}
\caption{\label{fig:res} Residuals of our fit of mean SEDs of main-sequence galaxies by the \citep{Draine2007} model.}
\end{figure*}

\begin{figure*}
\centering
\includegraphics{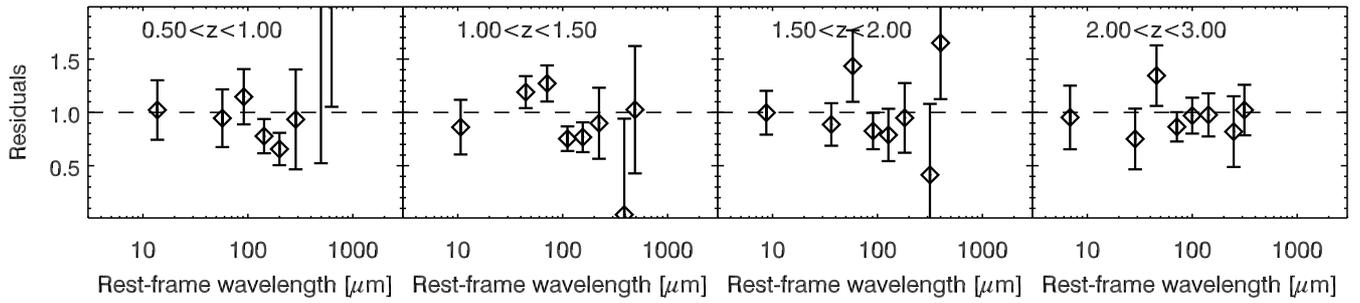}
\caption{\label{fig:res_sb10} Residuals of our fit of mean SEDs of strong starbursts by the \citep{Draine2007} model.}
\end{figure*}

Figures\,\ref{fig:res} and \ref{fig:res_sb10} shows the residuals of the fits of our mean SEDs derived by stacking. We did not find any systematic trend, except a 2$\sigma$ underestimation of the millimeter data in main-sequence galaxies at z$>$3.\\

\end{appendix}

\end{document}